\documentclass[aps,prd,amsmath,amssymb,reprint,showpacs,superscriptaddress,floatfix]{revtex4-2}
\usepackage{graphicx}
\usepackage{bm}
\usepackage{float}
\usepackage{bookmark}
\usepackage{comment}
\usepackage{feynmp-auto}
\usepackage{hyperref}
\hypersetup{colorlinks=true, linkcolor=blue, citecolor=blue, urlcolor=black}
\usepackage{xcolor}

\newcommand{\orcid}[1]{\href{https://orcid.org/#1}{\raisebox{-0.1\baselineskip}{\includegraphics[height=1.8ex]{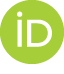}}}}

\begin{document}

\title{The Higgs Mechanism and Higgs Boson: Unveiling the Symmetry of the Universe}

\author{Ahmed Abokhalil\,\orcid{0000-0003-1400-2683}}
\affiliation{Department of Physics, Faculty Of Science, Zagazig University}

\date{June 1, 2023}

\begin{abstract}
The discovery of the Higgs boson at the Large Hadron Collider marked a significant milestone in particle physics. The Higgs mechanism, a key theoretical framework, provides profound insights into the origin of particle masses and the spontaneous breaking of symmetries. In this review article, we delve into the fundamental aspects of the Higgs mechanism and Higgs boson physics while exploring related topics such as Noether's theorem, gauge invariance, and the $U(1)$ and $SU(2)$ models. Additionally, we examine Goldstone's theorem, the extension of the Higgs mechanism to $U(1)$ and $SU(2)$ models, and its role in the electroweak theory for understanding the masses of the $W$ and $Z$ bosons. Finally, we scrutinize the properties of the Higgs boson itself, its production mechanisms, and decay processes, shedding light on its significance in unraveling the mysteries of the universe.

\end{abstract}

\maketitle

\section{Introduction}
Understanding the fundamental building blocks of matter and the forces governing their interactions has long been a quest at the forefront of physics. In the realm of particle physics, our understanding has experienced a revolution thanks to the discovery of the Higgs mechanism and the subsequent detection of the Higgs boson.

Proposed in the 1960s by François Englert, Robert Brout \cite{PhysRevLett.13.321}, Peter Higgs \cite{higgs1964broken}, Gerald Guralnik, C. R. Hagen, and Tom Kibble \cite{PhysRevLett.13.585}, the Higgs mechanism represents a profound breakthrough in comprehending how particles acquire mass. At its core lies the ethereal Higgs field, which permeates the entirety of space. Symmetry has always been a guiding principle in physics, and what makes the Higgs mechanism captivating is its involvement in the breaking of a specific symmetry. According to the theory, during the initial moments after the Big Bang, particles were devoid of mass, and all forces were unified. However, as the universe cooled and underwent a phase transition, the Higgs field emerged, spontaneously breaking the symmetry and giving birth to the diverse masses of particles observed today.

Particles interact with the Higgs field, acquiring mass as a consequence. Imagine particles moving through a medium—those that interact more intensely with the Higgs field encounter greater resistance, analogous to moving through a denser substance, and consequently acquire more mass. This interaction provides a mechanism for particles to obtain their distinct masses, offering an explanation for the varying weights of different particles.

The Higgs mechanism further predicts the existence of the Higgs boson, a particle that corresponds to the Higgs field. In a landmark achievement, the Higgs boson was eventually discovered in experiments conducted by ATLAS \cite{aad2012observation} and CMS \cite{chatrchyan2012observation} collaborations at CERN in 2012. This momentous detection confirmed the presence of the Higgs field and validated the theoretical framework proposed several decades earlier.

The discovery of the Higgs boson and the confirmation of the Higgs mechanism have profound implications for our comprehension of the universe. Not only have they provided a crucial puzzle piece in particle physics, but they have also shed light on the origins of mass and the underlying symmetries that shape our universe. Ongoing research and exploration in this field continue to deepen our understanding of the fundamental nature of matter and the forces that govern its behavior.

\section{\label{sec:level1}Noether's Theorem and Gauge Invariance}
Symmetry is an age-old idea that originated in geometry. It refers to the property of an object being unchanged by certain transformations, such as rotating a square around an axis that connects the centers of two parallel sides by multiples of 90 degrees, resulting in the square being in the same position and orientation. Geometry recognizes four principal kinds of symmetry: translation, rotation, reflection, and glide reflection. Although initially applied in geometry, symmetry can also be expanded to other fields. In physics, symmetry means that a given law or property remains unchanged (invariant) under certain transformations, such as spatial and time transformations. This property implies that we cannot physically differentiate between two distinct configurations of the dynamical variables that describe a system, meaning that the action remains invariant. For example, Newton's second law of motion, $\vec{F}=m\vec{a}$ is a mathematical equation that does not change when applied in different places and times, indicating that it has spatial and time symmetries and is invariant under these transformations.
Over a century ago, physicists realized that nearly all natural phenomena can be explained by assuming that nature seeks to optimize a certain function, known as the action, denoted by $S$. Rather than applying forces on masses to anticipate the progression of an event, which is determined by an initial condition (as in the case of Newton's laws of motion), we can begin with the boundary conditions that describe how the event starts and concludes and account for what happens in between by assuming that nature minimizes the action. Sometimes, nature maximizes the action rather than minimizing it. As a result, referring to the principle as the "least action principle" may not always be accurate. Instead, it is preferable to discuss the "principle of stationary action," in which a small variation in the system's behavior results in no change in the action.
\subsection{Linking symmetries and conservation laws}
In 1918, Emmy Noether \cite{noether1971invariant} discovered the relationship between the symmetries of action and conserved quantities.

The function $\mathcal{L}(\mathrm{\phi_a(\mathrm{x}), \partial_\mu \phi_a(\mathrm{x})})$ which depends on the filed $\phi(\mathrm{x})$ and its derivative $\partial_\mu\phi(\mathrm{x})$ is called the Lagrangian density. The action is defined as the integral of the Lagrangian density with respect to  spatial and temporal coordinates as:
\begin{equation*}S=\int d^4x\:\mathcal{L} .\end{equation*}

The principle of stationary action states that the path taken by the system between two points in space-time is the one that makes the action stationary, i.e., the variation in the action due to small variations in the path is zero. Mathematically, this is expressed as: 
\begin{equation*}\delta S=\delta \int d^4 x \mathcal{L}(\mathrm{\phi_a(\mathrm{x}), \partial_\mu \phi_a(\mathrm{x})})=0\end{equation*}
where $\delta S$ represents the variation in the action due to small variations in the path of the system.\\
The Euler-Lagrange equation is given by
\begin{equation*}\frac{\partial \mathcal{L}}{\partial \phi_a}-\partial_\mu \frac{\partial \mathcal{L}}{\partial(\partial_\mu \phi_a)}=0,\quad  (a=1,2,3,\ldots,n), \end{equation*}
where $\phi_a$ are the fields and $\mathcal{L}$ is the Lagrangian density.
The momentum field is defined as
\begin{equation*}\pi_a\mathrm(x)=\frac{\partial \mathcal{L}}{\partial(\partial_0 \phi_a)},\end{equation*}
and to quantize the fields, we introduce the following commutation relations:
\begin{equation*}\left[\phi_a(\mathrm{{\bf x},t}),\phi_b(\mathrm{{\bf y},t})\right]=\left[\pi_a(\mathrm{{\bf x},t}),\pi_b(\mathrm{{\bf y},t})\right]=0,\end{equation*}
\begin{equation*}\left[\phi_a(\mathrm{{\bf x},t}),\pi_b(\mathrm{{\bf y},t})\right]=i \delta_{ab}\delta^3({\bf x-y}).\end{equation*}

 The commutation relations between field operators give rise to fluctuations. These fluctuations can be thought of as quantum fluctuations or fluctuations of the underlying field \cite{peskin2018introduction}. They represent the inherent variability or uncertainty associated with the field's value at a given point in space and time.

These field fluctuations have important physical implications. They can give rise to the creation and annihilation of particles, and they affect the behavior of the field and its interactions. For example, the presence of quantum fluctuations in the electromagnetic field can lead to the spontaneous creation and annihilation of particle-antiparticle pairs, a phenomenon known as vacuum fluctuations.
If the transformation $\phi_a \rightarrow \phi_a + \alpha \Delta\phi_a$ is symmetric, wherein $\alpha$ represents an infinitesimal parameter and $\Delta \phi_a$ denotes a deformation in the field configuration, it follows that the Lagrangian must remain invariant under this transformation.
\begin{equation*}
    \mathcal{L}\rightarrow \mathcal{L}+ \alpha \:\partial_\mu \mathcal{J}^\mu ,
\end{equation*}
for some $\mathcal{J}^\mu$. The variation of the Lagrangian density, resulting from the variation of the field, is expressed as follows:
\begin{equation*}
\alpha \Delta \mathcal{L}=\frac{\partial \mathcal{L}}{\partial \phi_a}(\alpha \Delta \phi_a)+\left(\frac{\partial \mathcal{L}}{\partial(\partial_\mu \phi_a)}\right)\partial_\mu( \alpha \Delta \phi_a).\end{equation*}
We can use the product rule to represent the second term as 
\begin{align*}
&\alpha \: \partial_\mu\left(\frac{\partial \mathcal{L}}{\partial\left(\partial_\mu \phi_a\right)}\Delta \phi_a\right) - \alpha\: \partial_\mu \left(\frac{\partial\mathcal{L}}{\partial\left(\partial_\mu\phi_a\right)}\right)\Delta\phi_a \\
& = \alpha \:\partial_\mu\left(\frac{\partial \mathcal{L}}{\partial\left(\partial_\mu \phi_a\right)}\right)\Delta \phi_a \\
& \quad + \alpha \: \frac{\partial\mathcal{L}}{\partial\left(\partial_\mu\phi_a\right)}\:\partial_\mu \left(\Delta\phi_a\right) - \alpha  \:\partial_\mu\left(\frac{\partial\mathcal{L}}{\partial\left(\partial_\mu\phi_a\right)}\right)\Delta\phi_a \\
& = \left(\frac{\partial \mathcal{L}}{\partial(\partial_\mu \phi_a)}\right)\partial_\mu( \alpha \Delta \phi_a).
\end{align*}

After rearranging terms we get
\begin{equation*}\alpha \Delta \mathcal{L}=\alpha \left[\frac{\partial \mathcal{L}}{\partial \phi_a}-\partial_\mu\left(\frac{\partial \mathcal{L}}{\partial(\partial_\mu \phi_a)}\right)\right]\Delta \phi_a+\alpha \:\partial_\mu\left(\frac{\partial \mathcal{L}}{\partial(\partial_\mu \phi_a)}\Delta \phi_a\right).\end{equation*}
By the Euler-Lagrange equation, the term in the first parentheses vanishes, yielding
\begin{equation*}\alpha\Delta \mathcal{L}=\alpha \:\partial_\mu\left(\frac{\partial \mathcal{L}}{\partial(\partial_\mu \phi_a)}\Delta \phi_a\right).\end{equation*}
Setting the remaining term equal to $\alpha\:\partial_\mu \mathcal{J}^\mu$, we find that
\begin{equation*}
\partial_\mu j^\mu=0, \qquad \text{for} \qquad  j^\mu=\frac{\partial \mathcal{L}}{\partial(\partial_\mu \phi_a)}\Delta \phi_a - \mathcal{J}^\mu.\end{equation*} 
Then the current $j^\mu$ is conserved for continuous symmetries.

Separating the time differential and spatial differential of the current, we obtain
\begin{equation*}\frac{\partial}{\partial t} j^0+\vec{\nabla}\cdot \vec{j}=0.\end{equation*}
Integrating this equation over all space gives
\begin{equation*}\int d^3x\left(\frac{\partial}{\partial t} j^0+\vec{\nabla}\cdot \vec{j}\right)=0.\end{equation*}
As $\vec{j}$ vanishes at spatial infinity, we can use Gauss' theorem to write 
\begin{equation*}Q(\mathrm{t})=\int d^3x\:  j^0.\end{equation*}
Differentiating Gauss' theorem with respect to time gives
\begin{equation*}\frac{dQ(\mathrm{t})}{dt} =\frac{d}{dt}\int d^3x \:j^0,\end{equation*}
then we obtain
\begin{equation*}\frac{dQ(\mathrm{t})}{dt}=0.\end{equation*}

The fact that $\partial_\mu j^\mu = 0$ is known as the conservation law and it can be associated with a conserved quantity, which is given by the integral over space of the time component of the current $j^0$:
\begin{equation*}Q(\mathrm{t})=\int d^3x \: j^0(\mathrm{{\mathbf{x}},t}).\end{equation*}

If $\partial_\mu j^\mu = 0$, then $Q(t)$ is conserved, that is $\dot{Q}(\mathrm{t})=0$.
This is the core of Noether's theorem, which claims that a corresponding conserved quantity exists for any continuous symmetry of the Lagrangian. In this case, the charge $Q(t)$ corresponds to the conserved quantity, and the continuous symmetry is represented by the transformation $\phi_a(\mathrm{\mathbf{x},t}) \rightarrow \phi_a(\mathrm{\mathbf{x},t}) + \alpha \Delta \phi_a(\mathrm{\mathbf{x},t})$.
Because the Lagrangian density $\mathcal{L}$ is invariant under that transformation, the charge $Q(t)$ is conserved. This is obvious from the fact that the action $S = \int d^4x\, \mathcal{L}$ is also assumed to be invariant under the transformation. The charge $Q(t)$ is therefore a conserved quantity associated with this continuous symmetry according to Noether's theorem.

\subsection{The significance of gauge invariance in modern physics}
A theory is said to exhibit symmetry under a group $G$ if, when we apply transformations using the members of $G$, the fundamental components or entities of the theory (such as the variables describing the system) remain unchanged, and the Lagrangian, which describes the system's dynamics, remains invariant \cite{zee2016group}. These transformations can involve changes in space-time coordinates or changes in the fields.
In physics, there are two types of symmetries to consider. First, a theory can have a global symmetry, meaning that it remains unchanged under a transformation of the form $e^{i\theta}$, where $\theta$ is a constant but arbitrary phase. Global symmetries are associated with conserved quantities and can be explained using Noether's theorem.
\begin{equation*}\psi'(\mathrm{x})=U_\theta\psi(\mathrm{x})=e^{-i \theta}\psi(\mathrm{x}),\end{equation*}
where $\theta$ is constant in all spacetime. If the phase $\theta$ of the fermion field is not observable, then the Lagrangian is invariant under global transformation. It is obvious that $U_\theta^\dagger U_\theta =1$, and $U_{\theta_1} U_{\theta_2}=U_{\theta_2} U_{\theta_1}$, thus, this transformation is unitary and Abelian. This transformation is denoted by $U(1)$.

On the other hand, a theory can have a gauge symmetry, which asserts that the mathematical description of a physical theory should remain unchanged under specific transformations known as gauge transformations. These transformations are local, meaning they can vary from point to point in space and time.
\begin{equation*}\psi'(\mathrm{x})=e^{-i \theta(\mathrm{x})}\psi(\mathrm{x}),\end{equation*}
where the phase is $x-$dependent, and varies from point to point in space-time.
Gauge invariance holds immense significance in modern physics, as it serves as a foundational principle underlying our understanding of the fundamental forces and particles that govern the universe. This principle plays a crucial role in the development of gauge theories, such as quantum electrodynamics (QED), quantum chromodynamics (QCD), and the electroweak theory.

At its core, gauge invariance imposes constraints on the behavior of physical fields, requiring them to transform in a specific manner under these gauge transformations to maintain the invariance of the theory.

The significance of gauge invariance arises from its close connection to the fundamental forces of nature. In gauge theories, the gauge fields mediate the interactions between particles and carry the corresponding forces. Gauge invariance ensures the consistency and self-conservation of these theories by governing the behavior of the particles and fields involved.

One notable example of the significance of gauge invariance is found in quantum electrodynamics, the theory that describes the electromagnetic force. The gauge field in this theory is the electromagnetic vector potential, and gauge invariance is associated with the conservation of electric charge. The requirement of gauge invariance leads to the prediction of the photon as the force carrier for electromagnetic interactions. It provides a framework for understanding and calculating phenomena such as the interaction of charged particles, the behavior of electric and magnetic fields, and the propagation of electromagnetic waves.

Gauge invariance also holds immense importance in the electroweak theory, which unifies the electromagnetic and weak nuclear forces. The gauge invariance in this theory is associated with the $U(1)$ and $SU(2)$ gauge transformations. It ensures the self-consistency of the theory and gives rise to the prediction of the $W$ and $Z$ bosons as the force carriers for weak interactions. The Higgs mechanism, intimately connected to gauge invariance, plays a critical role in this theory by spontaneously breaking symmetry and providing masses to the $W$ and $Z$ bosons while preserving gauge invariance.

\section{\label{sec:level2}Local Gauge Symmetries and Gauge Fields}

At lower energy levels, the strong, weak, and electromagnetic interactions appear to be disconnected, as evidenced by their distinct coupling constants, which differ significantly from one another. However, in the realm of particle physics, there is a possibility that, under extraordinarily high energy conditions, the coupling constants associated with the strong, weak, and electromagnetic interactions may exhibit a tendency to converge toward a single, unified value. This postulates that all elementary particle forces are distinct manifestations of the same fundamental force, as proposed by Howard Georgi and Sheldon L. Glashow  \cite{georgi1974unity}. A significant breakthrough towards unification was achieved by Glashow, Weinberg, and Salam when they successfully merged the weak and electromagnetic interactions \cite{lim2004physics}. A crucial realization in this pursuit was that all fundamental interactions remain unchanged under local gauge transformations. It is hoped that gauge theories will serve as the foundation for a comprehensive unification of all fundamental interactions, representing a notable step in that direction.
The standard model of particle physics encompasses three distinct quantum gauge theories that elucidate the electromagnetic, weak, and strong interactions among elementary particles. These theories are renormalizable, meaning they can be mathematically adjusted to account for infinites and yield meaningful results. 
The $U(1)$ gauge theory, also known as quantum electrodynamics (QED), describes the electromagnetic force. It is based on the $U(1)$ symmetry group, where "$U$" stands for unitary and "1" denotes a one-dimensional representation. The gauge field in $U(1)$ theory is the electromagnetic vector potential, and the associated gauge transformations involve phase changes.

The $SU(2)$ gauge theory is a key component of the electroweak theory, which unifies the electromagnetic and weak nuclear forces. It is based on the $SU(2)$ symmetry group, where "$S$" denotes special unitary and "$2$" represents a two-dimensional representation. The gauge fields in $SU(2)$ theory are associated with the weak force and come in the form of a triplet of vector fields.
In $SU(2)$ gauge theory, the $W$ and $Z$ bosons are the force carriers responsible for weak interactions. The $SU(2)$ symmetry is related to the weak isospin, which characterizes the behavior of particles under weak interactions. 

The electroweak theory, incorporating both $U(1)$ and $SU(2)$ gauge theories, successfully describes phenomena such as beta decay, neutrino interactions, and the interaction of charged and neutral weak currents.

\subsection{\label{sec:level15} Quantum Electrodynamics and U(1) model}
Symmetry under group $U(1)$ in quantum electrodynamics (QED), refers to the fact that the equations describing the behavior of particles remain unchanged if the electron field and the gauge field are multiplied by a complex number of units of magnitude. These complex numbers can be represented by the exponential function $e^{i\theta}$, where $\theta$ is a real number between $0$ and $2\pi$. The group $U(1)$ is represented by a circle, where each point on the circle corresponds to an element of the group. Multiplying two elements in $U(1)$ involves adding their corresponding angles or parameters $\theta$, which corresponds to rotating around the circle. The phase of a wavefunction can be altered by multiplying it by a complex number. This corresponds to a change in the wavefunction's general "form" or "look," but not any of its actual characteristics, such as its probability distribution.

For instance, in quantum mechanics, the probability distribution of an atom's electron is determined by the electron's wavefunction, which reveals where the electron is most likely to be found. The probability distribution (which is the wavefunction's squared modulus) would not change if the wavefunction were multiplied by a complex number, but the wavefunction's overall shape would.

Similarly, in the framework of QED's $U(1)$ symmetry, multiplying the electron and gauge fields by complex numbers of unit magnitude alters their overall phase but has no effect on any of their physical characteristics or the theory's predictions.

Under local transformations, the Dirac Lagrangian undergoes a specific transformation that can be expressed mathematically as:
\begin{align*}
\mathcal{L}\rightarrow \mathcal{L}' &=\bar{\psi'}(\mathrm{x})\left(i\gamma^\mu\partial_\mu -m \right)\psi'(\mathrm{x})\\
&=e^{i \theta(\mathrm{x})}\bar{\psi}(\mathrm{x})\left(i\gamma^\mu\partial_\mu -m \right)e^{-i \theta(\mathrm{x})}\psi(\mathrm{x})\\
&=e^{i \theta(\mathrm{x})}\bar{\psi}(\mathrm{x})\gamma^\mu e^{-i \theta(\mathrm{x})}\psi(\mathrm{x})\partial_\mu \theta(\mathrm{x})\\
& \quad +e^{i \theta(\mathrm{x})}\bar{\psi}(\mathrm{x})e^{-i \theta(\mathrm{x})}\left(i\gamma^\mu\partial_\mu -m \right)\psi'(\mathrm{x})\\
&=\bar{\psi}(\mathrm{x})\left(i\gamma^\mu\partial_\mu -m \right)\psi'(\mathrm{x})+\bar{\psi}(\mathrm{x})\gamma^\mu \psi(\mathrm{x})\partial_\mu \theta(\mathrm{x})\\
&=\mathcal{L}+j^\mu (\mathrm{x})\partial_\mu \theta(\mathrm{x}),
\end{align*}
where $j^\mu =\bar{\psi}(\mathrm{x})\gamma^\mu \psi(\mathrm{x})$ is the vector current carried by the fermion. Thus, the Lagrangian is not invariant under local transformation unless $\partial_\mu \theta(\mathrm{x})=0$, which means that $\theta(\mathrm{x})$ is independent of $x$. The additional gradient-of-phase term spoils local phase invariance. Local phase invariance may be achieved, however, if the equations of motion and the observables involving derivatives are modified by the introduction of the electromagnetic field  $A_\mu $ as 
\begin{equation*}\partial_\mu \rightarrow D_\mu \equiv \partial_\mu - iq A_{\mu}(\mathrm{x}),\end{equation*}
then the Lagrangian becomes 
\begin{equation*}\mathcal{L}\rightarrow\mathcal{L}+q j^\mu A_\mu.\end{equation*}

Thus, if we define the Lagrangian by replacing $\partial_\mu$ by $D_\mu$
\begin{align*}
\mathcal{L}\left(\psi, \bar{\psi},A_\mu \right)&=\bar{\psi}\left(i\gamma^\mu D_\mu -m \right)\psi\\
&=\bar{\psi}\left(i\gamma^\mu\partial_\mu -m \right)\psi+q\bar{\psi}\gamma_\mu \psi A_\mu ,
\end{align*}
then this $\mathcal{L}$ becomes invariant under the transformation if at the same time we make the replacement 
\begin{equation}A_\mu \rightarrow A'_\mu =A_\mu -\frac{1}{q}\partial_\mu \theta(\mathrm{x}),\label{eq:1}\end{equation}
which precisely cancels the unwanted term. $D_\mu $ is called the covariant derivative.
Therefore, starting from an original Lagrangian which possesses a global symmetry and making the replacement Eq.\eqref{eq:1}, we get a new Lagrangian
\begin{equation}\mathcal{L}=\bar{\psi}\left(i\gamma^\mu D_\mu -m \right)\psi  \label{eq:2}\end{equation}
which is invariant under local gauge transformation by requiring 
\begin{equation*}\left(D_\mu \psi(\mathrm{x})\right)'=e^{-i \theta(\mathrm{x})}D_\mu \psi(\mathrm{x}).\end{equation*}

Then, electromagnetic dynamics is made invariant by introducing a spin $1$ vector (gauge) boson field $A_\mu $ called the photon through the covariant derivative, which is called "minimal coupling". It is very important to know that in the gauge invariant theories, the interaction between gauge bosons and particles (fermions and/or bosons) is uniquely determined only through the minimal coupling. The Lagrangian, which is invariant under local $U(1)$ gauge transformation, is therefore given by Eq.\eqref{eq:2}.
However, this is not the complete Lagrangian for describing the whole system. To arrive at the complete Lagrangian for quantum electrodynamics, it remains only to add a kinetic energy term for the vector field to describe the propagation of free photons and the mass term, which should be also a gauge invariant. 
For the mass term, the photon mass term would have the form 
\begin{equation*}\mathcal{L}_\gamma=\frac{m^2}{2}A_\mu A^\mu , \end{equation*}
which violates the local gauge invariance because
\begin{equation*}A^\mu A_\mu \rightarrow \left(A^\mu -\partial^\mu \theta\right)\left(A_\mu -\partial_\mu \theta\right)\neq A^\mu A_\mu.\end{equation*}

Therefore, to keep the Lagrangian invariant, we must set the mass of the photon to equal zero $m=0$. In the absence of sources, an appropriate Lagrangian for the free electromagnetic field is
\begin{equation*}\mathcal{L}=-\frac{1}{4}\left(\partial_\nu A_\mu -\partial_\mu A_\nu\right)\left(\partial^\nu A^\mu -\partial^\mu A^\nu\right),\end{equation*}
which can be written as 
\begin{equation*}\mathcal{L}=-\frac{1}{4}F_{\mu \nu}F^{\mu \nu}.\end{equation*}

Therefore, the complete gauge invariant Lagrangian for the system of an electron and a photon takes the following form,
\begin{equation*}\mathcal{L}_{QED}=\bar{\psi}(\mathrm{x})\left(i\gamma^\mu D_\mu -m \right)\psi(\mathrm{x})-\frac{1}{4}F_{\mu \nu}F^{\mu \nu}.\end{equation*}

\begin{figure}[htbp]
\begin{minipage}[b]{0.45\linewidth}
\begin{fmffile}{electron-electron-scattering1}
\begin{fmfgraph*}(100,100)
    \fmfleft{i1,i2}
    \fmfright{o1,o2}
    \fmf{fermion}{i1,v1}
    \fmf{fermion}{v1,i2}
    \fmf{photon,label=$\gamma$}{v1,v2}
    \fmf{fermion}{v2,o2}
    \fmf{fermion}{o1,v2}
    \fmfdot{v1,v2}
    \fmflabel{$f$}{i2}
    \fmflabel{$f$}{i1}
    \fmflabel{$f'$}{o1}
    \fmflabel{$f'$}{o2}
\end{fmfgraph*}
\end{fmffile}
\end{minipage}
\hfill
\begin{minipage}[b]{0.45\linewidth}
  \centering
  \begin{fmffile}{absorption}
    \begin{fmfgraph*}(100,100)
    \fmfleft{i1,i2}
    \fmfright{o1}
    \fmf{fermion}{i1,v1}
    \fmf{fermion}{v1,i2}
    \fmf{photon,label=$\gamma$}{v1,o1}
    \fmfdot{v1}
    \fmflabel{$f$}{i2}
    \fmflabel{$f$}{i1}
        \end{fmfgraph*}
  \end{fmffile}
\end{minipage}
\vspace{10pt}
\caption{On the right side of the figure, a fermion is shown emitting or absorbing a photon. On the left side, a two-step process is illustrated, where a fermion is scattered by another fermion. In this process, the first fermion emits a photon, which is then absorbed by the second fermion.}
\label{fig:scattering123}
\end{figure}
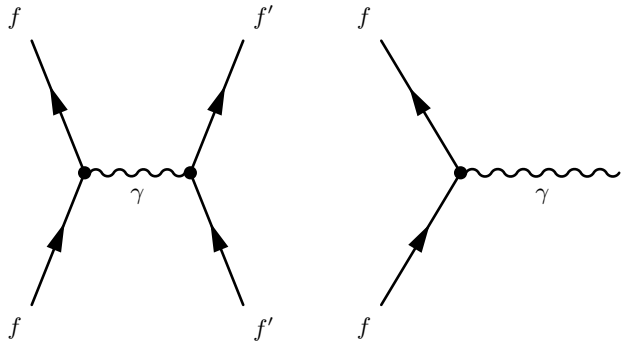

Quantum Electrodynamics (QED) is the field theory associated with the Lagrangian density $\mathcal{L}_{QED}$. It represents a generalization of classical electrodynamics to incorporate quantum phenomena. QED emerged as a result of efforts to describe the interactions between photons and electrons at the quantum level. In Quantum Electrodynamics (QED), the interactions between particles are mediated by massless photons. Photons are quanta of the electromagnetic fields, with no mass or charge, and possessing a spin of $1$. Importantly, photons do not engage in self-interactions. The theory of gauge fields refers to a collection of theories that build upon and go beyond Maxwell's electromagnetic field theory. There are two key components of Maxwell's theory. First, long-range behavior is seen in the forces created by gauge fields, i.e. photons, and obeying the inverse square law seen in Coulomb force. Second, the charge, a quantum feature of the source, which is directly related to the strength of the force, is conserved. Figure. \ref{fig:scattering123} shows common physical phenomena that occur through the influence of electromagnetic interactions. The principle of gauge invariance is widely acknowledged as the most influential guiding principle for comprehending not only Quantum Electrodynamics (QED) but also potentially all types of interactions.
\vspace{10pt}
\subsection{\label{sec:level3} Weak nuclear force and the SU(2) model}
Quantum Electrodynamics, which describes the electromagnetic interaction, was the first successful quantum field theory to incorporate gauge symmetries. QED provided a framework for understanding the behavior of electrically charged particles and the exchange of photons.

Building upon the success of QED, physicists sought to develop a unified theory that could encompass other fundamental interactions, such as the weak and strong nuclear forces. The aim was to find a mathematical framework that could incorporate gauge symmetries and describe all known particles and their interactions.

In the 1950s, Chen Ning Yang and Robert Mills \cite{PhysRev.96.191} proposed a generalization of gauge theories to include non-Abelian gauge symmetries. Yang-Mills theory provided a mathematical framework for describing interactions mediated by multiple gauge bosons, allowing for the incorporation of strong and weak nuclear forces.

The experimental evidence supporting the existence of strong and weak nuclear forces provided further motivation for the development of Yang-Mills theory. The observation of particle interactions that could not be explained solely by QED, such as beta decay and hadronic interactions, necessitated the inclusion of additional gauge symmetries and interaction carriers.

Consider two fermion fields, i.e. two spin $\frac{1}{2}$ fields, $\psi_1$ and $\psi_2$, then the Lagrangian of the system without any interactions is the sum of two Dirac Lagrangians:
\begin{align*}
\mathcal{L}&=\mathcal{L}_1+\mathcal{L}_2\\
&=\bar{\psi}_1(\mathrm{x})\left(i\gamma^\mu\partial_\mu -m_1 \right)\psi_1(\mathrm{x})\\
&\quad+\bar{\psi}_2(\mathrm{x})\left(i\gamma^\mu\partial_\mu -m_2 \right)\psi_2(\mathrm{x}).
\end{align*}
The Lagrangian can be written in more elegant way if we combine $\psi_1$, and $\psi_2$ into a column vector:
$$\psi=
\begin{pmatrix}
\psi_1 \\
\psi_2 \\
\end{pmatrix},
$$
and the adjoint spinor is:
$$\bar{\psi}=
\begin{pmatrix}
\bar{\psi_1}&\bar{\psi_2}
\end{pmatrix},
$$ 
and the mass matrix is:
$$M=
\begin{pmatrix}
m_1&0\\
0&m_2\\
\end{pmatrix},
$$
then the complete Lagrangian is:
\begin{equation}
\mathcal{L}=\bar{\psi}\left(i\gamma^\mu\partial_\mu -M\right)\psi . \label{eq:3} \end{equation}
Now, $\mathcal{L}$ can possess a broader global invariance compared to the previous case, as:
\begin{equation*} \psi\rightarrow \psi'= U\psi ,\end{equation*}
where $U$ is any $2\times 2$ unitary matrix.
thus, 
\begin{equation*}\bar{\psi}\rightarrow \psi'=\bar{\psi}U^\dagger .\end{equation*}
Therefore, $\bar{\psi}\psi$ is invariant.
As $U$ is unitary, it can be expressed as $U=e^{iH}$, where $H$ is Hermitian matrix. The Pauli matrices, a collection of three Hermitian $2\times 2$ matrices that serve as the basis for the space of $2\times 2$ Hermitian matrices, can be used to generate a general Hermitian matrix as:
\begin{equation*}H=a_0 I+a_1\tau_1+a_2\tau_2+a_3\tau_3=a_0 I+\bm{a}\cdot\bm{\tau}.\end{equation*}
Here, $a_0, a_1,a_2$, and $a_3$ are arbitrary coefficients, $I$ is the identity matrix, and $\tau_1,\tau_2$, and $\tau_3$ are Pauli matrices. Thus any unitary $2\times 2$ matrix can be expressed as:
\begin{equation*}U=e^{ia_0}e^{i\bm{a}\cdot\bm{\tau}}.\end{equation*}
We will concentrate on the transformation
\begin{equation*}\psi'=e^{i\bm{a}\cdot\bm{\tau}}\psi ,\end{equation*}
which is called global $SU(2)$ transformation. The Lagrangian Eq.\eqref{eq:3} is invariant under global $SU(2)$ gauge transformation. Yang and Mills extended this global invariance to a local gauge invariance. The first step is to make the parameter $a$ dependent on $x$, I will let $(\mathrm{x})=-\frac{1}{2}g\:\theta(\mathrm{x})$. Here,  $g$ is the coupling strength to be determined from experiments. Thus,
\begin{equation*}\psi'=e^{-ig\frac{\tau}{2}\theta(\mathrm{x})}\psi ,\end{equation*}
and its adjoint 
\begin{equation*}\bar{\psi'}=e^{ig\frac{\tau}{2}\theta(\mathrm{x})}\bar{\psi}.\end{equation*}
Then the gradient transforms as:
\begin{equation*}\partial_\mu \psi'=\left(\partial_\mu U \right)\psi + U\left(\partial_\mu \psi\right).\end{equation*}
Therefore, the new Lagrangian is:
\begin{align*}
\mathcal{L}'&=\bar{\psi}\left(i\gamma^\mu\partial_\mu -M \right)\psi\\
&=e^{ig\frac{\tau}{2}\theta(\mathrm{x})}\bar{\psi}\left(i\gamma^\mu\partial_\mu -M \right)e^{-ig\frac{\tau}{2}\theta(\mathrm{x})}\psi\\
&=\bar{\psi}\left(i\gamma^\mu\partial_\mu -M \right)\psi+\left(\partial_\mu \theta(\mathrm{x})\right)\bar{\psi}\left(\gamma^\mu g \frac{\tau}{2}\right) \psi\\
&=\mathcal{L}+\left(\partial_\mu \theta(\mathrm{x})\right)\bar{\psi}\left(\gamma^\mu g \frac{\tau}{2}\right) \psi.
\end{align*}

So, again the Lagrangian is not invariant under this transformation. To ensure the local gauge invariance of the theory, we first introduce a gauge covariant derivative,
\begin{equation*}\partial_\mu\rightarrow D_\mu=\partial_\mu -i g \vec{A}_\mu.\end{equation*}
The covariant derivative should satisfy
$$\left(D_\mu \psi\right)'=D_\mu '\psi'=U(D_\mu \psi)
$$
Thus,
\begin{align}
D_\mu '\psi'&=\left( \partial_\mu -i g \vec{A'}_\mu \right)\psi' \nonumber \\
&=\left(\partial_\mu U\right)\psi+U\left(\partial_\mu \psi\right)-igA'_\mu \left(U \psi\right),\label{eq:4}
\end{align}
also we have
\begin{align}
U(D_\mu \psi)&=U\left(\partial_\mu -i g \vec{A}_\mu\right)\psi \nonumber\\
&=U(\partial_\mu\psi)-igU(A_\mu \psi).\label{eq:5}
\end{align}
From Eq.\eqref{eq:4}, and Eq.\eqref{eq:5}, we conclude 
\begin{equation*}igA'_\mu \left(U \psi\right)=\left(\partial_\mu U\right)\psi+igU(A_\mu \psi),\end{equation*}
which must hold for arbitrary values of the field $\psi$. Inserting $I = U^{-1}U$ before each occurrence of $\psi$, we obtain
\begin{align*}
igA'_\mu \left(U U^{-1}U\psi\right)&=\left(\partial_\mu U\right)U^{-1}U\psi+igU(A_\mu U^{-1}U\psi)\\
igA'_\mu \psi'&=\left(\partial_\mu U\right)U^{-1}\psi'+igU(A_\mu U^{-1} \psi').
\end{align*}

\begin{widetext}
Regarding the transformation law as an operator
equation, we obtain
\begin{equation}A'_\mu =UA_\mu U^{-1}-\frac{i}{g}\left(\partial_\mu U\right)U^{-1} .\label{eq:6} \end{equation}
To construct the Lagrangian density for the dynamical part of
the gauge field in a gauge invariant manner, we demand that $F_{\mu \nu}$ transforms covariantly under gauge transformation.
$$F_{\mu \nu }\rightarrow UF_{\mu \nu}U^{-1}.
$$
We note that under the gauge transformation Eq.\eqref{eq:6}, the tensor representing the Abelian field strength would transform as

\begin{align*}
F_{\mu \nu}&=\partial_\mu A_\nu -\partial_\nu A_\mu\\
&\rightarrow\partial_\mu\left[UA_\nu U^{-1}-\frac{i}{g}\left(\partial_\nu U\right)U^{-1}\right] -\partial_\nu\left[UA_\mu U^{-1}-\frac{i}{g}\left(\partial_\mu U\right)U^{-1}\right]\\
&=\partial_\mu(UA_\nu U^{-1})-\frac{i}{g}(\partial_\mu \partial_\nu U)U^{-1}-\frac{i}{g}(\partial_\nu U)(\partial_\mu U^{-1})-\partial_\nu(UA_\mu U^{-1}) +\frac{i}{g}(\partial_\nu \partial_\mu U)U^{-1}+\frac{i}{g}(\partial_\mu U)(\partial_\nu U^{-1})\\
&=\frac{i}{g}\left(\partial_\mu U \partial_\nu U^{-1}-\partial_\nu U\partial_\mu U^{-1}\right)\partial_\mu U A_\nu U^{-1} +U\partial_\mu A_\nu U^{-1}+UA_\nu \partial_\mu U^{-1} -\partial_\nu U A_\mu U^{-1} -U\partial_\nu A_\mu U^{-1}-UA_\mu \partial_\nu U^{-1}\\
&=\frac{i}{g}\left(\partial_\mu U \partial_\nu U^{-1}-\partial_\nu U\partial_\mu U^{-1}\right) +\partial_\mu U A_\nu U^{-1} +U A_\nu\partial_\mu U^{-1}-\partial_\nu U A_\mu U^{-1}-UA_\mu \partial_\nu U^{-1}+U\left(\partial_\mu A_\nu-\partial_nu A_\mu\right)U^{-1}.\
\end{align*}

We see that unlike in the case of QED, here $F_{\mu\nu}$ is neither invariant nor does it have a simple transformation under the gauge transformation \cite{das2020lectures}. Let us also note that under the gauge transformation

\begin{align*}
&ig(A_\mu A_\nu - A_\nu A_\mu)\\
&\rightarrow ig\left[\left(U A_\mu U^{-1}-\frac{i}{g} (\partial_\mu U)U^{-1}\right)\left(UA_\nu U^{-1}-\frac{i}{g}(\partial_\nu U)U^{-1}\right)-\left(UA_\nu U^{-1}-\frac{i}{g}(\partial_\nu U)U^{-1}\right)\left(UA_\mu U^{-1}-\frac{i}{g}(\partial_\mu U)U^{-1}\right)\right]\\
&=ig\Big[U A_\mu A_\nu U^{-1}-\frac{i}{g}\left(\partial_\mu U A_\nu U^{-1} - U A_\mu \partial_\nu U^{-1}\right)+\frac{1}{g^2}\partial_\mu U \partial_\nu U^{-1} - U A_\nu A_\mu U^{-1}+\frac{i}{g}\left(\partial_\nu U A_\mu U^{-1} - U A_\nu \partial_\mu U^{-1}\right)\\
&\quad -\frac{1}{g^2} \partial_\nu U \partial_\mu U^{-1}\Big],
\end{align*}
\end{widetext}
thus, we can define the field tensor as 
\begin{equation*}F_{\mu \nu}=\partial_\mu A_\nu -\partial_\nu A_\mu-ig[A_\mu,A_\nu],\end{equation*}
in order to make it transform covariantly under the gauge transformation Eq.\eqref{eq:6}. It is now easy to construct the gauge invariant Lagrangian density for the dynamical part of the gauge field as
\begin{align*}
\mathcal{L}_{gauge}&=-\frac{1}{2}Tr\left(\vec{F}_{\mu \nu}\cdot\vec{F}^{\mu \nu}\right)\\
&=-\frac{1}{2}\sum_{i,j=1}^3 Tr\left(\frac{\tau^i}{2}F^{i}_{\mu \nu}\frac{\tau^j}{2}F^{j\mu \nu}\right)
= -\frac{1}{4}F^{i}_{\mu \nu}F^{i \mu \nu},\
\end{align*}
where the last equality comes from the relation $Tr\left(\frac{\tau^i}{2}\frac{\tau^j}{2}\right)=\frac{1}{2}\delta^{ij}$.
The Yang–Mills Lagrangian,
\begin{align*}
\mathcal{L}_{YM}&=\mathcal{L}_F+\mathcal{L}_{gauge}\\
&=\bar{\psi}\left(i\gamma^\mu D_\mu -M\right)\psi  -\frac{1}{4}F^{i}_{\mu \nu}F^{i \mu \nu},
\end{align*}
is, therefore, invariant under local gauge transformations. Whereas a mass term $m^2 A_\mu A^\mu$ is incompatible with local gauge invariance, so we set the mass to equal zero.
The Yang-Mills theory, a non-Abelian gauge theory, differs significantly from Abelian gauge theories like QED. In Yang-Mills theory, self-interactions among gauge fields prevent it from being a free theory even without matter fields. This contrasts with QED, where photons do not exhibit self-coupling. Unfortunately, the original form of Yang-Mills theory is not suitable for describing weak interactions. It predicts identical couplings to right- and left-handed fermions, resulting in parity conservation. To maintain gauge invariance, it is crucial to have massless gauge bosons (denoted as $A_\mu$) that would lead to weak interactions with an infinite range, similar to electromagnetism. However, this contradicts experimental observations. Only by introducing a mass term that breaks the gauge symmetry can the theory be reconciled with experimental results and achieve agreement with reality.

\section{\label{sec:level4} Goldstone's Theorem and the Higgs Mechanism}
\subsection{spontaneous symmetry breaking and goldstone's theorem}
Nature exhibits symmetries, some of which are precise, while others are only rough approximations. For instance, the Earth can be modeled as a perfect sphere for various purposes since it has rotational symmetry. Nevertheless, we are aware of deviations that only matter when we need to consider the Earth's gravitational field with great accuracy. Technically, the Earth's gravitational potential has a leading monopole term that represents its spherical symmetry, but there are also smaller dipole and quadrupole terms that indicate the breakdown of rotational symmetry.
There are two approaches to addressing symmetry breaking in field theory. One method involves introducing a term for symmetry breaking alongside the term for symmetry manually, as shown below:
\begin{equation*}\mathcal{L}=\mathcal{L}_{sym}+\mathcal{L}_{breaking}.\end{equation*}
This method is effective when the symmetry breaking term is small. In this situation, $\mathcal{L}$ retains the exact symmetry of a vanishing $\mathcal{L}_{breaking}$. However, it is worth noting that this circumstance is somewhat theoretical, since there is no fundamental principle to dictate the precise structure of $\mathcal{L}_{breaking}$.

Another way is called hidden or spontaneous symmetry breaking (SSB), where the Lagrangian remains symmetric under certain group transformations while the physical vacuum is made non-invariant. We denote the free part of the Klein-Gordon Lagrangian density for a complex scalar field as
\begin{equation*}\mathcal{L}_o=\partial_\mu \phi^\dagger \partial^\mu \phi -m^2\phi^\dagger \phi .\end{equation*}
To include interactions, the Lagrangian density for the interaction must be invariant under Lorentz transformations, translations, and gauge transformations if the theory has gauge symmetry. Furthermore, since the free Lagrangian density is invariant under the discrete $\mathbb{Z}_2$ symmetry transformation
\begin{equation*}\phi(x)\leftrightarrow \phi^\dagger(x),\end{equation*}
we would also like to preserve this symmetry in interactions. With these conditions, the simplest interaction Lagrangian density for a complex scalar field has the form
\begin{equation*}\mathcal{L}_I=-\frac{\lambda}{4}(\phi^\dagger \phi)^2 . \end{equation*}
This term describes the most general renormalizable self-interacting theory for a complex Klein-Gordon field.
The system is equivalent to the one described by the following Lagrangian composed of $2$ real fields $\varphi_1$ and $\varphi_2$ which are related to $\phi$ and $\phi^\dagger$ as $\phi=(\varphi_1+i\varphi_2)/\sqrt{2}$ and $\phi^\dagger=(\varphi_1-i\varphi_2)/\sqrt{2},$
\begin{equation*}\mathcal{L}=\frac{1}{2}\partial_\mu \varphi_1 \partial^\mu \varphi_1 +\frac{1}{2} \partial_\mu \varphi_2 \partial^\mu \varphi_2-V(\varphi_1^2+\varphi_2^2),\end{equation*}
which has $O(2)$ symmetry in two dimensions, meaning that rotating the field values at each point by an angle does not change the overall properties of the field,
$$\begin{pmatrix}
    \varphi_1\\
    \varphi_2
\end{pmatrix} \rightarrow \begin{pmatrix}
    \varphi_1'\\
    \varphi_2'
\end{pmatrix}=\begin{pmatrix}
    \cos\theta && -\sin\theta\\
    \sin\theta&& \cos\theta
\end{pmatrix}\begin{pmatrix}
    \varphi_1\\
    \varphi_2
\end{pmatrix}.$$

The potential $V$ is important to determine the behavior and stability of the system. To make the theory meaningful, $V$ must not have higher-order terms beyond the $4th$ order of the fields. This is related to the concept of renormalizability in quantum field theories, which ensures that the theory is physically meaningful. If the potential has higher order terms, the theory would be non-renormalizable and make meaningless predictions. Additionally, $V$ must be bounded below, which ensures that the system has a stable ground state with the lowest energy level. Without this condition, the system could have negative energy and unstable behavior.

In theories of quantum fields, particles that are created are considered to be excited states of the field. These excitations are quantized fluctuations of the field, which occur around its lowest energy state, or vacuum state. The value of the field at this state is referred to as the vacuum expectation value (VEV). To determine the spectra of particles, the potential is expanded around its minimum at the lowest energy state as
\begin{align*}
&V(\varphi_1,\varphi_2)=V(\varphi_{01},\varphi_{02})+\sum_{a=1,2}\left(\frac{\partial V}{\partial \varphi_a}\right)_0(\varphi_0-\varphi_{0a})\\
&\quad + \frac{1}{2}\sum_{a,b=1,2}\left(\frac{\partial^2 V}{\partial\varphi_a \partial \varphi_b}\right)_0(\varphi_a-\varphi_{0a})(\varphi_b-\varphi_{0b})+\cdots,
\end{align*}
where $\phi_0 = (\varphi_{01},\varphi_{02})$ is the (VEV) of $\phi = (\varphi_{1},\varphi_{2})$.
\begin{figure}
\begin{minipage}[b]{0.45\linewidth}
  \centering
  \includegraphics[width=\linewidth]{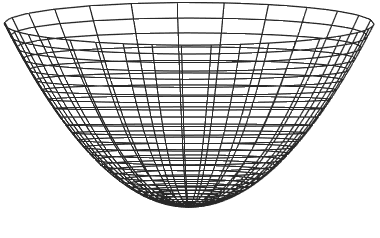}
  \caption{Potential with a unique minimum at $\varphi_1 = \varphi_2 = 0.$ }
  \label{fig:photo1}
\end{minipage}
\hspace{0.5cm}
\begin{minipage}[b]{0.45\linewidth}
  \centering
  \includegraphics[width=\linewidth]{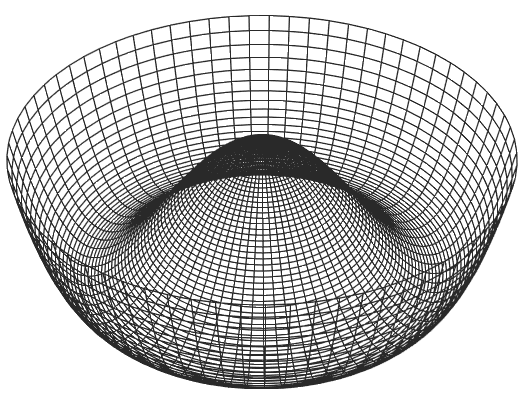}
  \caption{Potential leading to spontaneous breaking of symmetry.}
  \label{fig:photo2}
\end{minipage}
\end{figure}

Since the potential $V$ has its minimum at $\phi=\phi_0$, the 2nd term of the r.h.s. of the potential is zero. The factor $\frac{\partial^2 V}{\partial\varphi_a \partial \varphi_b}\equiv m_{ab}^2$ the 3rd term is called the mass matrix which is diagonalized to generate the particle spectrum. A positive value of the parameters $ m^2 > 0$ and $\lambda > 0,$ corresponds to the ordinary case of unbroken symmetry, illustrated by the potential in Fig. \ref{fig:photo1}. The unique minimum, corresponding to the vacuum state, occurs at $\varphi_{01}=\varphi_{02}=0$. The mass matrix becomes diagonal in this case
$$
m_{ab}^2 = 
\begin{pmatrix}
    m^2 && 0 \\
    0 && m^2
\end{pmatrix},
$$
which means that $\varphi_1$ and $\varphi_2$ have the same mass $m$.

Another one is the case where the vacuum is not unique and is called the "Nambu-Goldstone phase". This case is realized, for example, for continuously or infinitely degenerated vacuum states with $\varphi_{01}$ and/or $\varphi_{02}$ , as shown in Fig. \ref{fig:photo2}, for the potential with $m^2 =-\mu^2\quad (\mu^2 > 0)$ and $\lambda > 0$,
\begin{equation}V(\varphi_1^2+\varphi_2^2)=-\frac{\mu^2}{2}(\varphi_1^2+\varphi_2^2)+\frac{\lambda}{4}(\varphi_1^2+\varphi_2^2)^2 .\label{eq:7}\end{equation}
In general, the minimum of the potential is obtained by requiring
\begin{equation*}\left(\frac{\partial V}{\partial \varphi_1}\right)_0=-\mu^2\varphi_{01}+\lambda \varphi_{01}(\varphi_1^2+\varphi_2^2)=0,\end{equation*}
\begin{equation*}\left(\frac{\partial V}{\partial \varphi_2}\right)_0=-\mu^2\varphi_{02}+\lambda \varphi_{02}(\varphi_1^2+\varphi_2^2)=0,\end{equation*}
which leads to the condition
\begin{equation*}\varphi_{01}^2+\varphi_{02}^2=\frac{\mu^2}{\lambda}=v^2,\end{equation*}
or 
\begin{equation*}(\phi^\dagger \phi)_0=|\phi_0|^2=\frac{\mu^2}{2\lambda}=\frac{v^2}{2},\end{equation*}
where $v$ is new variable. In other words,  the minima lie on a circle of radius $ v = \sqrt{\mu^2/ \lambda}$ in the $(\varphi_1,\varphi_2)$ plane, that is, the vacuum state is no longer unique but is $O(2)$ symmetric. One can choose any point as the physical vacuum. From Eq.\eqref{eq:7}, we obtain 
\begin{align*}
    \frac{\partial^2 V}{\partial \varphi_1^2}&=-\mu^2+\lambda(\varphi_1^2+\varphi_2^2)+2\lambda \varphi_1^2 ,\\
    \frac{\partial^2 V}{\partial \varphi_2^2}&=-\mu^2+\lambda(\varphi_1^2+\varphi_2^2)+2\lambda \varphi_2^2 ,\\
    \frac{\partial^2 V}{\partial \varphi_1 \partial \varphi_2}&=2\lambda \varphi_1 \varphi_2 .
\end{align*}
We may as well pick $\varphi_{01}=v,\quad \varphi_{02}=0$ as the physical vacuum, we obtain the mass matrix as
$$m^2_{ab}=\begin{pmatrix}
    2\lambda v^2&&0\\
    0&&0
\end{pmatrix}. $$
We introduce new fields, $\eta$ and $\xi$, shifting the orginial field to the true vacuum
\begin{equation*} \eta(x)=\varphi_1(x)-\sqrt{\frac{\mu^2}{\lambda}}, \qquad \xi=\varphi_2 .\end{equation*}
Where $\eta(x)$ represents the quantum fluctuations about the minimum. Rewriting the Lagrangian in terms of these new field variables, we find 
\begin{align*}
    \mathcal{L}&=\left[\frac{1}{2}(\partial_\mu \eta)(\partial^\mu \eta)-\mu^2 \eta^2\right]+\left[\frac{1}{2} (\partial_\mu \xi)(\partial^\mu \xi)\right]\\
    &\quad -\lambda \sqrt{\frac{\mu^2}{\lambda}}\: \eta (\eta^2+\xi^2)-\frac{\lambda}{4}\left(\eta^2+\xi^2 \right)^2+\frac{\mu^4}{4\lambda}.
\end{align*}

The initial expression includes three terms. The first term refers to the free Klein-Gordon Lagrangian for the $\eta$ field with a massive mass of $m_\eta = \sqrt{2\mu^2}$. The second term refers to the free Lagrangian for the $\xi$ field, which has a mass of $m_\xi = 0$. The third and fourth terms consists of five couplings, and the last term is a constant that does not affect the outcome \cite{griffiths2020introduction}. The revised Lagrangian no longer has $O(2)$ symmetry, although the original Lagrangian explicitly exhibits it. This is because the vacuum symmetry was broken, leading to what is known as a hidden symmetry or spontaneous symmetry breaking (SSB). In this scenario, it is important to note that the $\xi$ field is naturally massless, and this is not a random occurrence. Goldstone's theorem \cite{goldstone1962broken} established that spontaneous breaking of a continuous global symmetry will always result in the appearance of spin-0 particles that are massless. These particles are commonly known as "Goldstone bosons." In general, the number of massless Goldstone bosons produced is dependent on the symmetry properties of the spontaneous symmetry breaking.

However Goldstone's theorem have serious problems in its predictions: first, it predicts a massless scalar (the $\xi$ field) particle (a Goldstone boson), however this prediction is not observed in nature. The absence of such a particle has been a significant challenge for the original Goldstone model.
The presence of a massless particle would imply certain consequences, such as long-range forces associated with the exchange of the Goldstone boson. These forces are not observed in nature, and experiments have not detected a particle corresponding to the Goldstone boson. Therefore, the absence of the observed massless scalar particle is considered problematic for the original Goldstone model. Second, the model described the dynamics of spontaneous symmetry breaking and the emergence of the Goldstone boson but did not account for its interactions with other particles, so it can’t describe the real world.

The Goldstone theorem itself remains a valid and important theoretical concept, but it is necessary to incorporate additional mechanisms.
\subsection{\label{sec:level5}The Higgs mechanism as a mechanism for mass generation }

In the previous subsection, we discussed the Goldstone theorem, which elucidates the concept of spontaneous symmetry breaking within the context of global symmetry. Moving forward, we can investigate the implications of a sudden disruption of local symmetry. To facilitate our analysis, we will revisit the self-interacting theory of a complex scalar field and incorporate an Abelian gauge field, resulting in a scalar QED. The difference between the Goldstone and Higgs models is simply that the latter includes interactions.
\subsection{\label{sec:level6} The U(1) model }
The field of scalar electrodynamics involves investigating the interaction between an Abelian gauge field and a complex scalar field represented by $\phi=\varphi_1+i\varphi_2$.A $U(1)$-invariant theory that describes the electrodynamics of charged scalars in the absence of spontaneous symmetry breaking is accompanied by a remarkably simple Lagrangian. This Lagrangian can be expressed as follows:
\begin{equation}\mathcal{L}=|D^\mu \phi|^2-\mu^2 |\phi|^2-\lambda(\phi^\dagger \phi)^2-\frac{1}{4}F_{\mu \nu}F^{\mu \nu},\label{eq:8}\end{equation}
with $\lambda >0$, and 
\begin{equation*}\phi =\frac{\varphi_1+i\varphi_2}{\sqrt{2}},\end{equation*}
is a complex scalar field, and as usual
\begin{equation*}D_\mu=\partial_\mu -iqA_\mu,\end{equation*}
\begin{equation*}F_{\mu\nu}=\partial_\mu A_\mu-\partial_\mu A_\nu.\end{equation*}
The Lagrangian Eq.\eqref{eq:8} is invariant under $U(1)$ rotations,
\begin{equation*}\phi(x)\rightarrow \phi'(x)=e^{-i \alpha}\phi(\mathrm{x}),\end{equation*}
and under the local gauge transformations
\begin{equation*}\phi(x)\rightarrow \phi'(x)=e^{-i \alpha(x)}\phi(\mathrm{x}),\end{equation*}
with the replacement 
\begin{equation*}A_\mu \rightarrow A'_\mu =A_\mu -\frac{1}{q}\partial_\mu \alpha(x).\end{equation*}

There are two cases, depending upon the parameters of the effective potential. For $\mu^2 > 0$, the potential has a unique minimum at $\phi = 0$, and the exact Lagrangian symmetry is preserved. The spectrum is simply that of an ordinary QED of charged scalars, with a single massless photon $A_\mu$ and two scalar particles, $\phi$ and $\phi^\dagger$, with a common mass $\mu$. The potential has a continuum of absolute minima, corresponding to a continuum of degenerate vacuum, at
\begin{equation*}|\phi_0|^2=\frac{\mu^2}{2|\lambda|}=\frac{v^2}{2}.\end{equation*}

To explore the spectrum, we shift the fields to rewrite the Lagrangian Eq.\eqref{eq:8} in terms of displacements from the physical vacuum. The latter may be chosen,
\begin{equation*}\langle\phi\rangle_0=\frac{v}{\sqrt{2}},\end{equation*}
where $v > 0$ is a real number. We then define shifted field as 
\begin{equation*}\phi'=\phi-\langle\phi\rangle_0 .\end{equation*}
Which is conveniently parameterized in terms of
\begin{align*}
    \phi(x)&=\frac{1}{\sqrt{2}}(v+\eta)e^{i\xi/v}\\
    &\approx \frac{v+\eta +i\xi}{\sqrt{2}}.
\end{align*}
Then the Lagrangian appropriate for the study of small oscillations is given by: \cite{quigg2013gauge}
\begin{align*}
    \mathcal{L}_{so}&=\frac{1}{2}\left[(\partial_\mu \eta)(\partial^\mu \eta)+2\mu^2 \eta^2\right]+\frac{1}{2}(\partial_\mu \xi)(\partial^\mu \xi)\\
    &\quad -\frac{1}{4}F_{\mu \nu }F^{\mu \nu}+qvA_\mu (\partial^\mu \xi)+\frac{q^2 v^2}{2}A_\mu A^\mu+\cdots.
\end{align*}

The $\eta$-field, which corresponds to radial oscillations, has a mass equal to $\sqrt{-2\mu^2}$. The gauge field $A_\mu$ appears to have acquired a mass, but is mixed in the penultimate term with the
seemingly massless field $\xi$.
An astute choice of gauge will make it easier to sort out the spectrum of the spontaneously broken theory. To this end, it is convenient to rewrite
\begin{equation}\phi(x)\rightarrow\phi'(x)=e^{i\xi(x)/v}\phi(x)=\frac{1}{\sqrt{2}}(v+\eta).\label{eq:9}\end{equation}
\begin{equation}A_\mu (x)\rightarrow B_\mu (x) =A_\mu (x) -\frac{1}{qv}\partial_\mu \xi(x), \label{eq:10}\end{equation}
which corresponds to the phase rotation on the scalar field, which is called the "unitary gauge".
Under this unitary gauge transformation, we have
\begin{equation}D_\mu \phi(x)\rightarrow D'_\mu \phi'(x)=(\partial_\mu-iq B_\mu)\frac{1}{\sqrt{2}}(v+\eta),\label{eq:11}\end{equation}
and 
\begin{equation}F_{\mu \nu}(A)\rightarrow F_{\mu \nu }(B)=\partial_\mu B_\nu -\partial_\nu B_\mu. \label{eq:12}\end{equation}
Here we can easily show $F_{\mu \nu }(B) = F_{\mu \nu }(A)$ by substituting Eq.\eqref{eq:10} into Eq.\eqref{eq:12}, that is, the field-strength tensor $F_{\mu \nu}$ is gauge invariant as it should be. Substituting
Eqs. \eqref{eq:9}, \eqref{eq:10}, \eqref{eq:11}, and \eqref{eq:12} into Eq.\eqref{eq:8}, one can rewrite the Lagrangian as follows;
\begin{align*}
    \mathcal{L}_{so}&=\frac{1}{2}|\partial_\mu \eta -iqB_\mu (v+\eta )|^2-\frac{\mu^2}{2}(v+\eta)^2-\frac{\lambda}{4}(v+\eta)^4\\
    &\quad-\frac{1}{4}F_{\mu \nu }(b)F^{\mu \nu}(B)\\ 
    &=\frac{1}{2}\left[(\partial_\mu \eta)( \partial^\mu \eta) -2\mu^2 \eta^2\right] -\frac{1}{4}F_{\mu \nu }(B)F^{\mu \nu }(B)\\
    &\quad +\frac{q^2v^2}{2}B_\mu B^\mu+\frac{1}{2}q^2B_\mu B^\mu \eta (\eta +2v)-\lambda v\eta^3-\frac{\lambda}{4}\eta^4.
\end{align*}

The choice of gauge is crucial in determining the degrees of freedom of a physical system, especially in the Higgs mechanism. By a specific gauge choice, the $\xi$-particle can be eliminated from the Lagrangian, and it becomes the longitudinal component of the massive vector field $B_\mu$, as can be seen in the gauge transformation. Before spontaneous symmetry breaking, the system had four degrees of freedom with two scalars $\phi$ and $\phi^\dagger$ and two transverse polarizations of the massless vector field $A_\mu$. After spontaneous symmetry breaking, the system is reduced to one massive scalar boson, $\eta$, with mass $m_\eta =\sqrt{2\mu^2}$, and one gauge vector boson, $B_\mu$, with three degrees of freedom of polarization. The Goldstone boson adds an extra degree of freedom, which is absorbed by the gauge field during the Higgs mechanism. When the Higgs mechanism is applied to the $U(1)$ model, the photon, which is the gauge boson responsible for the electromagnetic force, obtains an effective mass due to the breaking of the $U(1)$ symmetry by the Higgs field. The photon is initially massless and travels at the speed of light, but when the Higgs field is excited, acquiring a non-zero vacuum expectation value (VEV), it creates a background field that interacts with the photon, giving it an effective mass. This mass does not mean that the photon becomes a massive particle, but it behaves as if it has a mass in certain situations, such as interacting with other particles or passing through a medium.
\subsection{\label{sec:level7} The SU(2) model }
Let us now expand on the $U(1)$ model to create the non-Abelian SU(2) model. This involves using the complex doublet field
$$\phi=\begin{pmatrix}
\phi_1\\
\phi_2
\end{pmatrix}.
$$
With $SU(2)$ symmetry, the gauge invariant Lagrangian is given by
\begin{equation*}\mathcal{L}=(D_\mu \phi)^\dagger(D^\mu \phi)-\frac{1}{4}F_{\mu \nu }^iF^{i\mu \nu}-V(\phi^\dagger \phi),\end{equation*}
which includes the terms
\begin{align*}
F_{\mu \nu }^i &= \partial_\mu A_\nu^i-\partial_\nu A_\mu^i+g\varepsilon_{ijk}A_\mu^j A_\nu^k ,\\
D_\mu \phi &= (\partial_\mu -ig\frac{\tau^i}{2}A_\mu^i)\phi ,\quad &&(i=1,2,3)\end{align*}
with the potential
\begin{equation*}
V(\phi^\dagger \phi) = -\mu^2\phi^\dagger \phi +\lambda (\phi^\dagger \phi)^2 . \qquad (\mu^2>0)
\end{equation*}

This Lagrangian ensures $SU(2)$ symmetry and is gauge invariant. We introduce new real fields, denoted by $H(x)$ and $\xi^i(x)$ (where $i$ ranges from 1 to 3). The field $\phi(x)$ is then expressed as a function of these new fields, and a parameterization is provided.
\begin{equation*}
\phi(x)=\frac{1}{\sqrt{2}}e^{i\tau^i \xi^i(x)/2v}\begin{pmatrix}
    0\\
    v+H(x)
\end{pmatrix}.
\end{equation*}

By taking the same unitary gauge as before, we define the new field $\phi'(x)$ as a function of the new fields.
$$\phi(x)\rightarrow \phi'(x)=U(x)\phi(x)=\frac{1}{\sqrt{2}}\begin{pmatrix}
    0\\
    v+H(x)
\end{pmatrix}.$$
We also introduce a new field $\vec{B}_\mu$ that is defined in terms of the original gauge field $\vec{A}_\mu$ and $U(x)$, 
\begin{equation*}\vec{A}_\mu\rightarrow\vec{B}_\mu=U\vec{A}_\mu U^{-1}-\frac{i}{g}(\partial_\mu U)U^{-1},\end{equation*}
with
\begin{equation*}U(x)=\frac{1}{\sqrt{2}}e^{i\tau^i \xi^i(x)/2v},\end{equation*}
where the summation over $i$ is implied.  Finally, a transformation is performed on the covariant derivative of $\phi(x)$, which is denoted as $(D_\mu \phi)'$, and expressed in terms of the new fields and the original gauge field $\vec{A}_\mu$. 
$$D_\mu \phi \rightarrow (D_\mu \phi)'=(\partial_\mu -ig \frac{\tau^i}{2}B^i_\mu)\frac{1}{\sqrt{2}}\begin{pmatrix}
    0\\
    v+H(x)
\end{pmatrix},$$
\begin{equation*}F_{\mu \nu }^i(A)F^{i\mu \nu}(A)\rightarrow F_{\mu \nu }^i(B)F^{i\mu \nu}(B)=F_{\mu \nu }^i(A)F^{i\mu \nu}(A),\end{equation*}
where the field tensor is given by:
\begin{equation*}F^i_{\mu \nu }(B)=\partial_\mu B_\nu^i-\partial_\nu B_\mu^i +g\varepsilon_{i j k } B_\mu^j B_\nu^k .\end{equation*}
The resulting Lagrangian is given by
\begin{equation*}\mathcal{L}=(D_\mu \phi)'^{\dagger}(D^\mu \phi)'-\frac{1}{4}F_{\mu \nu }^iF^{i\mu \nu}+\mu^2\phi'^{\dagger} \phi' -\lambda (\phi'{^\dagger} \phi')^2 .\end{equation*}

The Lagrangian derived shows that the three fields $\xi(x), (i = 1 ,2,3)$ are no longer present. It is unclear where these fields went, but we can determine their fate by expressing the Lagrangian in terms of the individual fields of $\phi'$. To do so, let's write the covariant derivative term first as,
\begin{widetext}
\begin{equation*}
[(D_\mu \phi)']^{\dagger a}(D^\mu \phi)'_{a}=\frac{1}{2}\partial_\mu H\partial^\mu H + g^2 B_\mu^i B^{j \mu }\left(\frac{\tau^i}{2}\right)_b^a\left(\frac{\tau^j}{2}\right)^c_a b'^{b} b'_c =\frac{1}{2}\partial_\mu H\partial^\mu H+g^2 B_\mu^i B^{j \mu } (v+H)^2.
\end{equation*}
Then, we finally obtain the following Lagrangian,
\begin{equation}
\mathcal{L} = \frac{1}{2}\left(\partial_\mu H\partial^\mu H -2\mu^2 H^2\right)- \frac{1}{4}F_{\mu \nu }^i(B) F^{i\mu \nu}(B) + \frac{g^2v^2}{8}B_\mu^i B^{i\mu} + \frac{g^2}{8}B_\mu^i B^{i \mu } H(2v+H) - \lambda vH^3 - \frac{\lambda}{4}H^4 - \frac{v^4}{4}.\label{eq:13}
\end{equation}
\end{widetext}

Thus, by choosing a specific gauge, the $\xi$ field is removed from the Lagrangian and becomes the longitudinal component of the massive vector field $B^i_\mu$ with $i$ ranging from $1$ to $3$, each with mass $m_B=\frac{1}{2}gv$. The Higgs boson $H$ also emerges with mass $m_H=\sqrt{2\mu^2}$, as evident from the Lagrangian Eq.\eqref{eq:13}. This symmetry-breaking process in $SU(2)$ theory is responsible for the emergence of the weak nuclear force, one of the four fundamental forces of nature. In the electroweak sector of the Standard Model, the Higgs field breaks the electroweak symmetry and gives rise to three massive gauge bosons (the $W^+$, $W^-$, and $Z^0$ bosons) and a massless gauge boson, the photon. The Higgs boson couples to both gauge bosons and fermions, and its mass determines the strength of these interactions. The Higgs mechanism is significant because it provides a means for particles to acquire mass without violating gauge invariance, a fundamental principle in particle physics. Experimental measurements of the $W^\pm$ and $Z$ boson masses support the predictions of the Standard Model.

\section{\label{sec:level8}Electroweak Theory and W, Z Boson Masses}

In 1967, Steven Weinberg \cite{weinberg1967model} and Abdus Salam \cite{salam1959weak} applied the Higgs mechanism \cite{higgs1964broken, PhysRevLett.13.321,PhysRevLett.13.585} to Sheldon Lee Glashow's \cite{GLASHOW1961579} electroweak unified theory, which marked the beginning of the Glashow-Weinberg-Salam (GWS) model of the Standard Model of particle physics. The GWS model unifies two of the four known fundamental forces of nature, the electromagnetic and weak forces, at an energy of $v = 246\:GeV$ and is described by four gauge bosons: the photon, which is the massless gauge boson of the electromagnetic interaction, and the $W^+, W^-$, and $Z$ bosons, which act as force carriers for the weak interaction. The Higgs mechanism is necessary to consistently attribute masses to the $W^\pm$ and $Z$ bosons, which allows the construction of a renormalizable theory, as was first demonstrated by Gerardus ’t Hooft and Martinus Veltman \cite{osti_4691149} in 1972. The GWS model is described by the $SU(2)_L \times U(1)_Y$ gauge symmetry.
Beta decay process could be interpreted as a neutron decaying into a proton and an electron. Pauli then postulated an additional massless, neutral particle ( the neutrino ) to ensure energy-momentum conservation. This novel object would be a participant only in the weak (nuclear) interaction, which must be the cause of beta decay are well-described by a specific form of the Fermi theory \cite{fermi}, which involves four-fermion interactions in the form of a current-current interaction Hamiltonian density:
\begin{equation*}\mathcal{H}_I=\frac{G_F}{\sqrt{2}}J^\dagger_\mu J^\mu .\end{equation*}
\vspace{10 pt}
\begin{figure}[!h]
\centering
\begin{fmffile}{neutron_decay}
\begin{fmfgraph*}(150,100)
  \fmfleft{i}
  \fmfright{o1,o2,o3}
  \fmf{fermion}{i,v}
  \fmf{fermion}{v,o1}
  \fmf{fermion}{v,o2}
  \fmf{fermion}{o3,v}
  \fmflabel{$n$}{i}
  \fmflabel{$p$}{o1}
  \fmflabel{$e^-$}{o2}
  \fmflabel{$\bar{\nu}_e$}{o3}
  \fmfdot{v}
\end{fmfgraph*}
\end{fmffile}
\vspace{10pt}
\caption{Feynman diagram for neutron decay using the feynmp-auto package. The neutron ($n$) decays into a proton ($p$), an electron ($e^-$), and an antineutrino ($\bar{\nu}_e$).}
\label{fig:neutron_decay}
\end{figure}
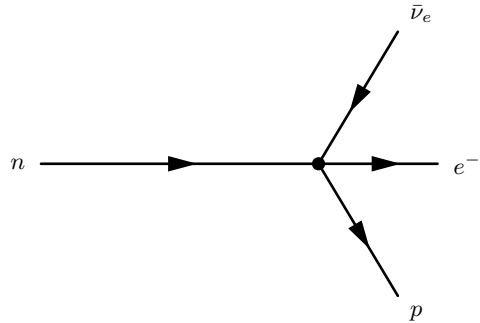
The combination of hadronic and leptonic currents is represented by the current $J_\mu$, while the Fermi coupling constant denoted by $G_F$ is derived from low energy experiments and equals $1.17 \times 10^{-5}(GeV)^{-2}$. Therefore, weak interactions are recognized for their limited range and weakness. Fermi successfully made predictions about the energy distribution of beta particles. Initially, his graphs were qualitative, but they were later found to match experimental measurements very accurately \cite{kurie1936radiations}. While Fermi's theory initially described the weak interaction, the discovery of parity violation indicated the necessity for a new approach. This had been proposed as early as 1956 by Yang and Lee \cite{lee1956question}, and was soon proved in experiments by Wu et al \cite{wu1957experimental} and by Goldhaber et al \cite{goldhaber1958helicity}.

The weak force affects particles such as leptons and hadrons, these particles belong to doublets that can be described by the weak isospin group of $SU(2)_L$, where the symmetry transformation only affects the left-handed components of the fields. Weakly interacting particles have an additive $U(1)$ quantum number called the weak hypercharge. Although weak interactions violate both weak isospin and weak hypercharge quantum numbers, they conserve electric charge in any weak process. The short range of the weak force suggests that massive gauge bosons are involved in a gauge theoretic description. Gauge bosons can acquire mass through the Higgs mechanism, which involves the spontaneous breakdown of some local symmetry. As weak interactions violate both the weak isospin and weak hypercharge symmetries, the Higgs mechanism can be associated with the spontaneous breakdown of these symmetries. Therefore, it is evident that a gauge theory of weak interactions should be sought after, which is based on the local symmetry group $SU(2)_L\times U(1)_Y$. Here, $SU(2)_L$ and $U(1)_Y$ represent the weak isospin and weak hypercharge symmetry groups, respectively. It is desirable that this symmetry group spontaneously breaks down at the weak interaction scale in such a way that only the electromagnetic symmetry is left as the true symmetry of the low-energy theory. This desired gauge theory should demonstrate the symmetry behavior:
\begin{equation*}SU(2)_L \times U(1)_Y \xrightarrow{\text{SSB}} U(1)_{EM}.\end{equation*}

The Glashow-Weinberg-Salam (GWS) model utilizes the $SU(2)_L \times U(1)_Y$ gauge theory and introduces a left-handed doublet and a right-handed singlet of the $SU(2)$ group as matter fields \cite{2023arXiv230407559R}. This choice is motivated by the observed properties of particles and the desire to incorporate both the weak force and electromagnetic force into a unified theory. 

The reason for this distinction between left-handed and right-handed components lies in the behavior of particles under the weak force. The weak force only interacts with left-handed particles and right-handed antiparticles. This is known as the chiral nature of the weak force.

By introducing the left-handed doublet (L) and the right-handed singlet (R), the electroweak theory allows for a consistent description of the weak force's interactions with particles. The left-handed components of the doublet participate in weak interactions, while the right-handed singlets do not.

The inclusion of both left-handed and right-handed components in the electroweak theory is necessary to account for the experimental observations and maintain the symmetry of the theory.

For example the fields that represent an electron and its neutrino, respectively, with respect to the $SU(2)_L \times U(1)_Y$ symmetry, can be expressed as:
$$L=\begin{pmatrix}
    \nu_e\\
    e
\end{pmatrix}_L, \quad R=e_R ,$$
where the right-handed component of electron $e_R$ has no interactions with any other
particles and thus it should be a singlet, and we assume that $\nu_e$ to be massless and thus $\nu_e$ has no right-handed component.
The process of building the model begins with the creation of a Lagrangian using the $L$ and $R$ that are invariant under the direct product of the $SU(2)_L$ and $U(1)_Y$ groups. To achieve this, a transformation of $L$ and $R$ is performed using group parameters for weak isospin and weak hypercharge operators, denoted by $\alpha^i (i = 1,2,3)$ and $\beta$ respectively. The $SU(2)_L$ and $U(1)_Y$ groups act on $L$ and $R$ differently, with $L$ transforming under both groups and $R$ only under $U(1)_Y$. 
\begin{align*}
SU(2)_L&: L\rightarrow L'=e^{-i\alpha^i(x)\frac{\tau^i}{2} }L, \quad R\rightarrow R'=R,\\
U(1)_Y&: L\rightarrow L'=e^{\frac{i}{2}\beta(x)}L, \quad R\rightarrow R'=e^{i\beta(x)}R.
\end{align*}

As the Lagrangian is locally gauge invariant, the group parameters $\alpha^i$ and $\beta$ depend on the space-time coordinate $x$.

The next step involves constructing a gauge invariant Lagrangian for fermions that maintains the symmetry of $SU(2)_L \times U(1)_Y$. This is achieved by using covariant derivatives for $L$ and $R$ in an equation that includes gauge boson fields associated with $SU(2)_L$ and $U(1)_Y$, denoted by $A_\mu^i$ (i = 1,2,3) and $B_\mu$ respectively. The gauge invariant Lagrangian for fermions can be expressed through the following equation:
\begin{align}
    \mathcal{L}_F&=\bar L i\gamma^\mu \left(\partial_\mu -i g\frac{\vec{\tau}}{2}\cdot \vec A_\mu +\frac{i}{2} g' B_\mu \right)L\nonumber\\
    &\quad + \bar R i\gamma^\mu \left(\partial_\mu +ig' B_\mu \right)R .\label{eq:bb}
\end{align}
The constants $g$ and $g'$ are used as the gauge couplings for $SU(2)_L$ and $U(1)_Y$, respectively. To obtain the covariant derivatives for $L$ and $R$, the general formula is used, which is given by
\begin{equation*}D_\mu =\partial_\mu -ig \frac{\vec{\tau}}{2} \cdot \vec A_\mu -ig' \frac{Y}{2} B_\mu,\end{equation*}
where $Y$ is equal to $-1$ for $L$ and $-2$ for $R$. Since $R$ is a singlet of $SU(2)_L$, it is not associated with $A_\mu^i$. At this stage, all fermions such as an electron and its neutrino, are massless because the fermion mass term, which links $L$ and $R$ fields, does not exist in Eq. \eqref{eq:bb}. The reason for this is that it validates $SU(2)_L \times U(1)_Y$ invariance. 
The kinetic term for the gauge fields, which must be included in the Lagrangian, is given by
\begin{equation*}\mathcal{L}_G=-\frac{1}{4}F^i_{\mu \nu}F^{i \mu \nu}-\frac{1}{4}B_{\mu \nu }B^{\mu \nu },\end{equation*}
with
\begin{align*}
F_{\mu \nu }^i &= \partial_\mu A_\nu^i-\partial_\nu A_\mu^i+g\varepsilon_{ijk}A_\mu^j A_\nu^k,\\
B_{\mu \nu }&=\partial_\mu B_\nu -\partial_\nu B_\mu,
\end{align*}
where $F_{\mu \nu }^i ( i = 1,2,3)$ and $B_{\mu \nu }$ are field strength tensors of gauge fields corresponding to $SU(2)_L$ and $U(1)_Y$, respectively. The mass terms for these gauge bosons do not appear due to local gauge invariance.
To achieve the necessary symmetry breaking, the Higgs mechanism employs scalar fields i.e., Higgs bosons. The process begins with four gauge bosons, three of which are connected with $SU(2)_L$, while one is associated with $U(1)_Y$. 

The aim is to end up with only one massless photon related to $U(1)_{EM}$. This can be accomplished by utilizing scalars that have at least four degrees of freedom. The minimal model of such scalars is the simplest example, which involves an $SU(2)$ doublet comprising two complex scalar fields. The doublet has a weak hypercharge of $Y_\phi = +1$ and is represented as:
\begin{equation}
\phi=\begin{pmatrix}
    \varphi^+\\
    \varphi^0
\end{pmatrix}. \label{eq:layla}
\end{equation}
Here, $\varphi^+$ and $\varphi^0$ refer to the positively charged and neutral complex scalar fields, respectively. The Lagrangian describing these scalars is defined by the equation:
\begin{equation*}\mathcal{L}_s=(D_\mu \phi)^\dagger (D^\mu \phi )-V(\phi^\dagger \phi).\end{equation*}
Here, $D_\mu \phi$ is given by:
\begin{equation*}D_\mu \phi =\left( \partial_\mu -ig \frac{\vec{\tau}}{2} \cdot \vec A_\mu -\frac{i}{2}g' B_\mu \right)\phi .\end{equation*}
The specific form of the covariant derivative is a result of $Y_\phi= +1$. The potential term $V(\phi^\dagger \phi)$ is gauge invariant and is represented as:
\begin{equation}V(\phi^\dagger \phi)=m^2\phi^\dagger \phi +\lambda(\phi^\dagger \phi)^2 .\label{eq:zz}\end{equation}
The parameters $m^2$ and $\lambda$ are both real constants, with $\lambda$ needing to be positive to ensure a stable vacuum. The theory must be renormalizable, which means that higher power terms of $\phi^\dagger \phi$ are not allowed.

We can incorporate Yukawa interaction terms between fermions and scalars, which are gauge invariant with respect to $SU(2)_L \times U(1)_Y$, in order to provide electrons with mass after spontaneous symmetry breakdown. These coupling terms are defined by:
\begin{equation*}\mathcal{L}_Y=-G_e \left(\bar L \phi R+\bar R \phi^\dagger L\right)+h.c.,\end{equation*}
where $G_e$ is the Yukawa coupling constant \cite{yukawa1935interaction}, which cannot be determined from the GWS model alone. The gauge invariance of $\mathcal{L}_Y$ with respect to $SU(2)_L \times U(1)_Y$ can be confirmed using the previously defined hypercharge values of $L$, $R$, and $\phi$. The full set of gauge invariant Lagrangians for the GWS model is the sum of the individual components presented above:
\begin{equation*}\mathcal{L}=\mathcal{L}_F+\mathcal{L}_G+\mathcal{L}_s+\mathcal{L}_Y .\end{equation*}
As before, spontaneous symmetry breaking occurs when the scalar doublet $\phi$ develops a vacuum expectation value
\begin{equation}
\phi_0=\langle 0|\phi|0\rangle =\begin{pmatrix}
    0\\
    v/\sqrt{2}
\end{pmatrix}. \label{eq:cc}
\end{equation}
Now, it is convenient to parametrize the scalar doublet with $4$ degrees of freedom in terms of the fields denoting the shifts from the vacuum state $\phi_0$,
$$
\phi=\begin{pmatrix}
    \varphi^+\\
    \varphi^0
\end{pmatrix}=e^{i\vec{\tau}\cdot \vec{\xi}/2v}\begin{pmatrix}
    0\\
    (v+H)/\sqrt{2}
\end{pmatrix}.
$$
Here the original $2$ complex scalar fields $\varphi^+$ and $\varphi^0$ in Eq. \eqref{eq:layla} are replaced by $4$ real fields, $\xi_i(i = 1,2,3)$ and $H$, where $\xi_i$ are Goldstone bosons being absorbed into the longitudinal components of $W^\pm$ and $Z^0$ bosons as described later and $H$ is a Higgs boson. Eq. \eqref{eq:cc}  leads to zero vacuum expectation values for all of these fields
\begin{equation*}\langle 0|\xi_i|0\rangle =\langle0|H|0\rangle=0 .\end{equation*}
Here we can rewrite the Lagrangian in the "unitary gauge", where 3 Goldstone bosons disappear by being 'eaten up' by gauge bosons, $ W^\pm$, and $Z^0$, and thus, physical particle spectra and their interactions become apparent. By applying the unitary $SU(2)$ transformation
\begin{equation*}U(\xi)=e^{-i \vec\tau \cdot \vec \xi /2v},\end{equation*}
one can come to the real world induced in the unitary gauge. Then, we can define the new fields in our real world as
\begin{align*}
    \phi'&=U(\xi)\phi =\begin{pmatrix}
    0\\
    (v+H)/\sqrt{2}
\end{pmatrix}=\frac{1}{\sqrt{2}}(v+H)\chi,\\
L'&=U(\xi)L,\\
\vec A'_\mu&=U(\xi) \vec A_\mu U(\xi)^{-1} -\frac{i}{g}(\partial_\mu U(\xi))U^\dagger(\xi),
\end{align*}
with $\chi=\begin{pmatrix}
    0\\
    1
\end{pmatrix}$ and $\vec {\bf A}_\mu =\vec A_\mu \cdot \frac{\vec \tau}{2}$, where the new fields transformed from the
original ones are presented with a prime. $R$ and $B_\mu$ remain unchanged under this $SU(2)$ transformation,
\begin{align*}
    R'&=R,\\
    B'_\mu&=B_\mu.
\end{align*}
The Lagrangian is invariant under this transformation and one can rewrite each piece as
\begin{align*}
\mathcal{L}_F&=\bar L' i\gamma^\mu (\partial_\mu -ig\frac{\vec \tau}{2}\cdot \vec A'_\mu +\frac{i}{2}g'B'_\mu )L'\\
&\quad+\vec R' i\gamma^\mu (\partial_\mu +ig'B'_\mu )R',\\
\mathcal{L}_G&=-\frac{1}{4}F^{'i}_{\mu \nu}F^{'i\mu \nu }-\frac{1}{4}B'_{\mu \nu }B^{'\mu \nu},\\
\mathcal{L}_s&=(D_\mu \phi)'(D^\mu \phi)'-V(\phi^{'\dagger}\phi'),\\
\mathcal{L}_Y&=-G_e(\bar L' \phi' R'+\bar R'\phi^{'\dagger}L')+h.c.
\end{align*}
Now let us discuss the physics described by this Lagrangian realized in the
unitary gauge. First we consider the scalar sector. The scalar fields generate masses of gauge bosons and those of quarks and leptons via the Higgs mechanism. $\mathcal{L}_s$ is explicitly written as
\begin{equation}\mathcal{L}_s=(D_\mu \phi)'(D^\mu \phi)'-V(\phi^{'\dagger}\phi'),\label{eq:qq}\end{equation}
with 
\begin{align*}
    (D_\mu \phi)'&=(\partial_\mu -ig\frac{\vec \tau}{2}\cdot \vec A'_\mu -\frac{i}{2}g'B'_\mu )\phi'\\
    &=(\partial_\mu -ig\frac{\vec \tau}{2}\cdot \vec A'_\mu -\frac{i}{2}g'B'_\mu )\frac{1}{\sqrt{2}}(v+H)\chi .
\end{align*}
The first term of Eq.\eqref{eq:qq} contains the mass-squared term for weak gauge bosons which is originated from the quadratic terms of gauge fields as shown in the following,
\begin{align*}
    \mathcal{L}_{mass}&=\frac{v^2}{2}\chi^\dagger \left(g\frac{\vec\tau}{2} \cdot \vec A'_\mu +\frac{g'}{2}B'_\mu\right)\left(g\frac{\vec \tau }{2} \cdot \vec A'^{\mu}+\frac{g'}{2}B'^{\mu}\right)\chi\\
    &=\frac{v^2}{8}\left(g^2 \vec A'_\mu \cdot \vec A'^{\mu}+g'^2 B'_\mu B'^\mu -2gg'B'_\mu A'^{3\mu}\right)\\
    &=\frac{v^2}{8}\left(g^2  A'^1_\mu A'^{1\mu}+g^2  A'^2_\mu A'^{2\mu}+\left(g A'^3_\mu -g'B'_\mu\right)^2\right),
\end{align*}
where in turning from the 1st line to the 2nd line, we used the formula $\tau^i\tau^j = \delta_{ij} + i\varepsilon^{ijk}\tau^k$. Now let us introduce charged boson fields $W^\pm$ defined by
\begin{equation*}W^\pm_\mu=\frac{A'^1_\mu \mp i A'^2_\mu}{\sqrt{2}}.\end{equation*}
Then the sum of the 1st and 2nd terms of  $\mathcal{L}_{mass}$ can be written as $\frac{1}{4}g^2 v^2 W^+_\mu W^{-\mu}$. It means that the charged vector bosons $W^\pm$ are massive with the mass
\begin{equation*}M_W=\frac{1}{2}gv .\end{equation*}
The remaining term which is described by neutral fields can be written as
$$\frac{v^2}{8}\begin{pmatrix}
    A'^3_\mu&& B'_\mu
\end{pmatrix}\begin{pmatrix}
    g^2&&-gg'\\
    -gg'&&g'^2
\end{pmatrix}\begin{pmatrix}
    A'^{3\mu}\\
    B'^\mu
\end{pmatrix},$$
which can be diagonalized into
\begin{equation}
\frac{v^2}{8}\begin{pmatrix}
    Z_\mu&&A_\mu
\end{pmatrix}\begin{pmatrix}
    g^2+g'^2&&0\\
    0&&0
\end{pmatrix}\begin{pmatrix}
    Z^\mu\\
    A_\mu 
\end{pmatrix}=\frac{v^2}{8}(g^2+g'^2)Z_\mu Z^\mu,\label{eq:yy}\end{equation}
by an orthogonal transformation
$$\begin{pmatrix}
    Z_\mu\\
    A_\mu 
\end{pmatrix}=\begin{pmatrix}
    \cos\theta_W&& -\sin\theta_W\\
    \sin\theta_W&&\cos\theta_W
\end{pmatrix}\begin{pmatrix}
    A'^3_\mu\\
    B'_\mu
\end{pmatrix},$$
where $\theta_W$ is called the weak mixing angle or Weinberg angle. The diagonalization leads to
\begin{equation*}\tan \theta_W=\frac{g'}{g},\end{equation*}
or 
\begin{equation*}\sin \theta_W=\frac{g'}{\sqrt{g^2+g'^2}},\qquad \sin\theta_W=\frac{g}{\sqrt{g^2+g'^2}}.\end{equation*}
From Eq.\eqref{eq:yy}, we see that the neutral $Z$ boson becomes massive with the mass
\begin{equation}
M_Z=\frac{1}{2}v\sqrt{g^2+g'^2},\label{eq:brbr}
\end{equation}
and another neutral boson $A_\mu$ is massless and hence can be identified with the real photon. Note that in the GWS model the mass of $Z^0$ boson is related to the one of $W^\pm$ bosons as
\begin{equation}
M_z=\frac{M_W}{\cos\theta_w}. \label{eq:krkr}
\end{equation}

One can see that the masses of $W^\pm$ and $Z^0$ are quite large, and can be determined in term of experimentally well known quantities. The parameter $v$ in Eq. \eqref{eq:brbr}  and Eq.\eqref{eq:krkr} can be expressed in terms of $G_F$ as $v = (G_F \sqrt{2})^{-1/2}$. And parameter $g$ and $g'$ can be expressed in terms of electric charge and weak mixing angle as $g \sin \theta_W = g' \cos\theta_W = e$.  we can estimate the masses to be
\begin{equation*}M_W\approx \frac{37.22}{\sin\theta_W}\:GeV>37\:GeV ,\end{equation*}
\begin{equation*}M_Z=\frac{M_W}{\cos\theta_W}\approx\frac{74.44}{\sin 2\theta_W}\:GeV>74\:GeV .\end{equation*}

The values of $M_W$ and $M_Z$ are obtained if $\sin \theta_W$ is determined experimentally. Actually, the value of $\sin^2 \theta_W$ is obtained experimentally \cite{sirunyan2018measurement} to be around $0.23101 \pm 0.00053$, leading to $M_W \approx 77.43 GeV$ and $M_Z \approx 88.3GeV$. In our previous derivation of the masses of the $W$ and $Z$ bosons, we overlooked the effects of radiative corrections. Accounting for these corrections necessitates a discussion of renormalization, which is beyond the scope of this paper. Therefore, I will simply present the final result for the renormalized (or physical) masses:
\begin{equation*}
    M_W = 79.8 \pm 0.8\:GeV,\qquad M_Z = 90.8 \pm 0.6\:GeV .
\end{equation*}
These values are in good agreement with the experimental masses \cite{cdf2022high}:
\begin{equation*}
   M_W= 80.4335\pm 0.094\:GeV,\quad M_Z=91.1876\pm 0.021 \:GeV .
\end{equation*}
The potential term Eq.\eqref{eq:zz} of scalars becomes, after symmetry breaking,
\begin{align}
    V(\phi'^\dagger \phi')&=-\frac{\mu^2}{2}(v+H)^2\chi^\dagger \chi +\frac{\lambda}{4}(v+H)^4(\chi^\dagger \chi)^2\nonumber\\
    &=\frac{-\mu^2 v^2}{4}+\frac{1}{2}(2\mu^2)H^2+\lambda v H^3+\frac{\lambda}{4}H^4.\label{eq:kk}
\end{align}

 Eq.\eqref{eq:kk} reveals that the mass of the physical Higgs boson $H$ can be equated to
\begin{equation*}M_H=\sqrt{2\mu^2},\end{equation*}

However, within the framework of the Glashow-Weinberg-Salam (GWS) model, the value of the mass of Higgs boson cannot be predicted through any established principle. We proceed to examine the Lagrangian $\mathcal{L}_s$ under the conditions of the unitary gauge. By doing so, we can determine the resulting expression of the Lagrangian $\mathcal{L}_s$ while disregarding a constant term.
\begin{widetext}
\begin{align*}
\mathcal{L}_s&=(D_\mu \phi)'(D^\mu \phi)'-V(\phi'^\dagger \phi')\\
&=\frac{1}{2}\partial_\mu H \partial^\mu H-\frac{1}{2}  M_H^2 H^2-\lambda v H^3 -\frac{\lambda}{4}H^4+\frac{g^2}{8}(H^2+2Hv)\left[\frac{1}{\cos^2\theta_W }Z_\mu Z^\mu +2W_\mu^+W^{-\mu }\right]+M_W^2 W_\mu^+W^{-\mu }+\frac{1
}{2}M_Z^2 Z_\mu Z^\mu .
\end{align*}

Next, we will examine the Yukawa interaction term $\mathcal{L}_Y$ in the unitary gauge,

\begin{equation*}
    \mathcal{L}_Y=-G_e(\bar L' \phi' R'+\bar R'\phi^{'\dagger}L')+h.c.=-G_e\left(\bar e'_L \frac{1}{\sqrt{2}}(v+H)e'_R+\bar e'_R\frac{1}{\sqrt{2}}(v+H)e'_L\right)+h.c.=-\frac{G_e v}{\sqrt{2}}\bar e' e'-\frac{G_e}{\sqrt{2}}H \bar e' e'.
\end{equation*}
\end{widetext}

As the unitary gauge considers only the physical degrees of freedom, we can identify $e'$ as the physical electron and subsequently eliminate the prime $'$ notation. Yukawa interaction term involves the interaction between fermions (such as electrons or quarks) and the Higgs field. This interaction is responsible for giving mass to these fermions through the Higgs mechanism. Yukawa coupling constant $G_e$ determines the strength of the interaction and consequently the mass of the fermion. The initial term in this equation represents the electron's mass term with a mass equal to
\begin{equation*}m_e=\frac{G_e}{\sqrt{2}}v .\end{equation*}

A notable point to consider is that the mass of the electron is influenced by the vacuum expectation value ($v$) of the Higgs field. This relationship also holds true for the masses of weak gauge bosons. The second term shows an interaction term of an electron to the Higgs boson with the coupling constant
\begin{equation*}\frac{G_e}{\sqrt{2}}=\frac{m_e}{v},\end{equation*}
here the coupling constant is proportional to the electron mass. The larger the Yukawa coupling constant $G_e$, the stronger the interaction between the fermion and the Higgs field, resulting in a higher mass for the fermion.
\section{\label{sec:level9} Higgs Boson: Production and Decay}
\subsection{Overview of the Higgs boson discovery}
All matter, including ourselves and the world around us, is composed of tiny particles. However, during the early stages of the universe, no particles had mass; they all sped around at the speed of light. It was only when particles acquired mass through an essential field associated with the Higgs boson that the formation of stars, planets, and life as we know it became possible. The confirmation of this field's existence came in 2012 with the discovery of the Higgs boson particle by the ATLAS \cite{aad2012observation} and CMS \cite{chatrchyan2012observation} collaborations at CERN. The ATLAS experiment reported a mass of approximately $126 \: GeV$, while the CMS experiment reported a value of $125.3 \: GeV$ for the Higgs boson.

Introducing a new field in the theory implies the existence of a corresponding particle, as every particle can be understood as a wave in a quantum field. The predicted properties of this particle align with the theory, providing strong evidence supporting the BEH mechanism. Without the discovery of such a particle, we would lack the means to investigate the existence of the Higgs field. With a mass of more than $120$ times that of the proton, the Higgs boson is the second-heaviest particle known today. This large mass, combined with an extremely short lifetime $(10^{-22}\: \text{seconds})$ means that the particle cannot be found in Nature – its existence can only be verified by producing it in the lab.
\subsection{Experimental methods and detection techniques}
In the context of Higgs boson production, the principles of energy conservation dictate that the total energy of colliding particles must be at least twice the mass of the Higgs boson. This means that to create a Higgs boson with a mass of $125 \:GeV$, a minimum collision energy of around $250 \:GeV$ is required. Initially, the Large Hadron Collider (LHC) operated at collision energies of $7 \:TeV$, which were indeed sufficient to directly produce the Higgs boson.

When particles collide, the conservation of energy and momentum is crucial in the production of the Higgs boson. The collision energy determines the amount of energy available to form the Higgs boson and other particles that may be generated during the collision.

While it is true that the direct production of the Higgs boson can occur through parton-level processes, such as gluon-gluon fusion, which do not require extremely high collision energies, increasing the collision energy provides several advantages. Higher energies enhance the likelihood of Higgs boson production, leading to a larger number of events and improved statistical significance. This, in turn, enhances the accuracy of measurements and strengthens the confirmation of the Higgs boson's existence.

Partons, including quarks and gluons, play a crucial role in the production of the Higgs boson \cite{halzen2008quark}. Gluon-gluon fusion is the dominant mechanism for Higgs boson production at the LHC, where two gluons from the colliding protons interact to create a virtual Higgs boson, which subsequently materializes as a real Higgs boson. Other parton-level processes, such as quark-antiquark annihilation or quark-gluon scattering, also contribute to Higgs boson production, albeit to a lesser extent.

Understanding the distribution and interactions of partons within the colliding protons is vital for accurately predicting the rate and properties of Higgs boson production. Theoretical models and simulations incorporate this knowledge to interpret experimental data obtained from particle collider experiments like the LHC. This comprehensive understanding helps scientists analyze and interpret the results, further advancing our knowledge of the Higgs boson and its fundamental properties.

The detection of the Higgs boson, despite its production not being a challenge, presents a significant research endeavor due to its fleeting existence and subsequent decay into various particles. With a mass of around $125 \:GeV$, the Higgs boson stands as one of the heavier elementary particles known. Its remarkably short lifetime, on the order of $10^{-22}$ seconds, poses a substantial obstacle to direct observation. However, scientists have successfully employed indirect methods by studying the particles into which the Higgs boson decays. Among these decay channels, the decay to two photons $(\gamma \gamma)$ has been a primary focus. Despite its low branching ratio of approximately $0.2\:\% $, this channel offers a distinct decay signature with minimal quantum noise. Researchers at the ATLAS and CMS experiments conducted analyses of proton-proton collision data at $\sqrt{s} = 7$ and $8\: TeV$, enabling the exploration of the Higgs boson's production and decay rates across various channels, including the $\gamma \gamma$ channel. These measurements have played a pivotal role in determining the Higgs boson's properties, such as its mass and coupling strengths to other particles, furthering our understanding of fundamental particles and their interactions.
\vspace{10pt}
\begin{figure}[!h]
\centering
\begin{fmffile}{LHC}
\begin{fmfgraph*}(160,130)
  \fmfleft{i1,i2}
  \fmfright{o1,o2}
  \fmf{fermion}{i1,v1}
  \fmf{fermion}{i2,v2}
  \fmf{gluon,label=$g$}{v1,v3}
  \fmf{gluon,label.side=left,label=$g$}{v2,v4}
  \fmf{fermion,label=$t/b$}{v4,v3}
  \fmf{fermion}{v3,v5}
  \fmf{fermion}{v5,v4}
  \fmf{dashes,label=$H$}{v5,v6}
  \fmf{fermion}{v6,v8}
  \fmf{fermion}{v8,v7}
  \fmf{fermion}{v7,v6}
  \fmf{photon}{v7,o1}
  \fmf{photon}{v8,o2}
  \fmflabel{$\gamma$}{o1}
  \fmflabel{$\gamma$}{o2}
  \fmflabel{$P$}{i1}
  \fmflabel{$P$}{i2}
  \fmffreeze
  \fmfshift{10left}{v3,v4}
  \fmfshift{7left}{i2,v1}
  \fmfshift{7left}{i1,v1}
\end{fmfgraph*}
\end{fmffile}
\caption{Feynman diagrams showing the cleanest channels associated with the low-mass $(~125 GeV/c^2)$ Higgs boson candidate observed by ATLAS and CMS at the LHC. The dominant production mechanism at this mass involves two gluons from each proton fusing to a Top-quark Loop, which couples strongly to the Higgs field to produce a Higgs boson.}
\label{fig:LHC}
\end{figure}
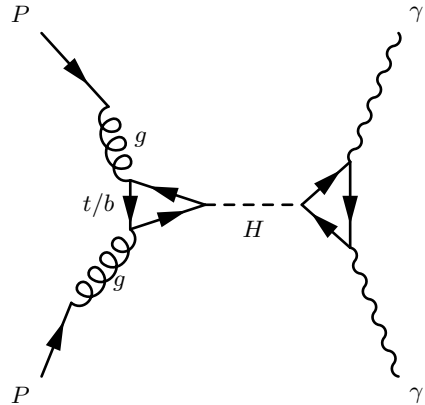
\subsection{Production mechanisms at colliders}
Producing the new particle is only the first step, however. Given its lifetime, the Higgs boson almost immediately decays – or transforms – into other particles. So it is not possible to observe it directly. The particles from the boson’s decay are the only traces that it leaves behind. These traces have to be detected and precisely measured by particle detectors. Once the decay products have been detected, the next step is to determine whether we can say that the Higgs boson was produced. The problem is that the particles that the Higgs decays into are the same kinds of particles that are copiously produced in particle collisions. Simply seeing a pair of photons (one of the final states from the Higgs boson decay) is hardly any indication that the Higgs boson exists and is being produced in the experiment. Especially since the Higgs boson is only produced about once in a billion of these collisions.  Scientists thus need some way of determining when a pair of photons (or four muons or a different final state that the Higgs decays into) is coming from a Higgs boson decay and when it’s not.

The production of Higgs bosons involves the acceleration of a large quantity of particles to extremely high energies, nearly reaching the speed of light. Subsequently, these particles are made to collide with each other. At the Large Hadron Collider (LHC), protons and lead ions (the nuclei of lead atoms) are utilized for this purpose. During these high-energy collisions, the desired rare particles, including the Higgs boson, can be generated and subsequently detected and studied. Any deviations or discrepancies from theoretical predictions can contribute to the improvement of the underlying theory, such as the Standard Model. The Standard Model provides guidance on the specific types of collisions and detectors required for Higgs boson production. According to the Standard Model, various processes can lead to the formation of Higgs bosons, although the likelihood of producing a Higgs boson in any given collision is expected to be very low. The most commonly expected methods for Higgs boson production are: 

{\bf Gluon fusion}

When hadrons like protons or antiprotons collide, as observed in particle accelerators such as the LHC and Tevatron, it is probable that two of the gluons responsible for binding the hadrons together will collide. The most straightforward way to generate a Higgs boson is through the combination of these two gluons, forming a loop of virtual quarks. Since the interaction between particles and the Higgs boson is proportional to their mass, this process is more likely to occur with heavy particles. In practice, it is sufficient to consider the contributions of virtual top and bottom quarks, which are the heaviest quarks. This process, involving virtual quarks, is the primary mechanism at play at the LHC and Tevatron, being approximately ten times more likely than any other processes. Therefore, it dominates the production of Higgs bosons in these accelerators.
\vspace{10pt}
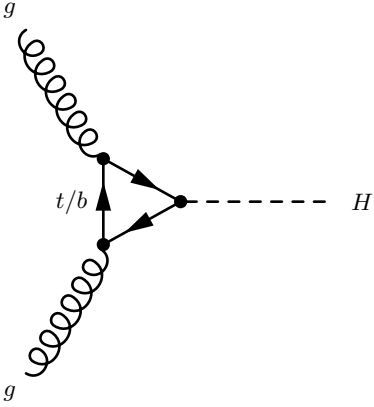
\begin{figure}[!h]
\begin{center}
\begin{fmffile}{gluon-gluons}
\begin{fmfgraph*}(130,130)
  \fmfleft{i1,i2}
  \fmfright{o1}
  \fmf{gluon}{i1,v1}
  \fmf{gluon}{i2,v2}
  \fmf{fermion,label=$t/b$}{v1,v2}
  \fmf{fermion}{v2,v3}
   \fmf{fermion}{v3,v1}
  \fmf{dashes}{v3,o1}
  \fmfdot{v1,v2,v3}
  \fmflabel{$H$}{o1}
  \fmflabel{$g$}{i2}
  \fmflabel{$g$}{i1}
\end{fmfgraph*}
\end{fmffile}
\end{center}
\caption{Feynman diagram depicting the interaction between gluons and the Higgs boson.}
\end{figure}

{\bf Higgs Strahlung}

When an elementary fermion encounters its corresponding anti-fermion, such as a quark with an anti-quark or an electron with a positron, they can combine to form a virtual $W$ or $Z$ boson. If this virtual boson carries sufficient energy, it has the potential to emit a Higgs boson. This particular process played a significant role in the production of Higgs bosons at the LEP, where the collision of an electron and a positron resulted in the formation of a virtual $Z$ boson. It also made a substantial contribution to Higgs production at the Tevatron. However, at the LHC, which collides protons with protons, this process ranks third in terms of significance. This is because the likelihood of a quark-antiquark collision is lower compared to the Tevatron. This phenomenon is referred to as Higgs Strahlung or associated production.
\vspace{10pt}
\begin{figure}[!h]
\centering
\begin{fmffile}{higgs-strahlung}
\begin{fmfgraph*}(120,120)
  \fmfleft{i1,i2}
  \fmfright{o1,o2}
  \fmf{fermion}{i1,v1}
  \fmf{fermion}{v1,i2}
  \fmf{photon,label=$W/Z$}{v1,v2}
  \fmf{dashes}{v2,o1}
  \fmf{photon}{v2,o2}
  \fmflabel{$H$}{o1}
  \fmflabel{$\bar{f}$}{i2}
  \fmflabel{$f$}{i1}
  \fmflabel{W/Z}{o2}
\end{fmfgraph*}
\end{fmffile}
\caption{Feynman diagram depicting the Higgs Strahlung process.}
\end{figure}
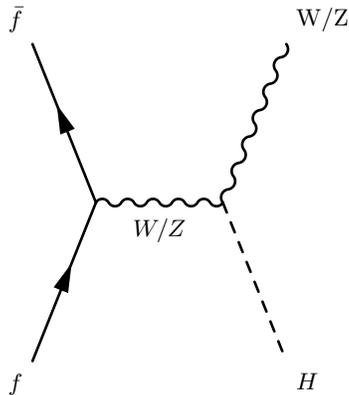

{\bf Weak boson fusion}

In the context of fermion collisions, there exists another intriguing possibility: the exchange of a virtual $W$ or $Z$ boson between two (anti-)fermions, leading to the emission of a Higgs boson. Notably, the colliding fermions are not constrained to be of the same type, allowing for diverse combinations. For instance, an up quark may engage in a $Z$ boson exchange with an anti-down quark. This particular process holds great significance as the second most prominent contributor to Higgs particle production at both the LHC and LEP experiments.
\vspace{10pt}
\begin{figure}[!h]
\centering
\begin{fmffile}{WeakBosonFusion}
\begin{fmfgraph*}(150,150)
  \fmfleft{i1,i2}
  \fmfright{o1,o2,o3}
  \fmf{fermion}{i1,v1}
  \fmf{fermion}{i2,v3}
  \fmf{fermion}{v1,o1}
  \fmf{fermion}{v3,o3}
  \fmf{photon,label=$W/Z$}{v1,v2}
  \fmf{photon,label=$W/Z$}{v3,v2}
  \fmf{dashes}{v2,o2}
  \fmflabel{$H$}{o2}
  \fmflabel{$f$}{i2}
  \fmflabel{$f$}{i1}
\end{fmfgraph*}
\end{fmffile}
\caption{Feynman diagram depicting the Weak Boson Fusion process.}
\label{fig:weak-boson-fusion}
\end{figure}
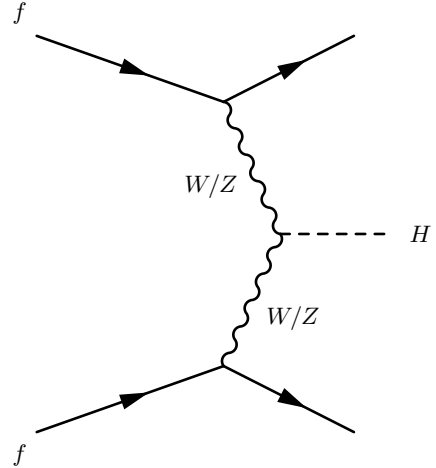

{\bf Top Fusion}

Among the various processes under consideration, the final one stands out as significantly less probable, with a lower likelihood by two orders of magnitude. This particular process entails the collision of two gluons, both of which undergo decay to produce a pair of heavy quarks and antiquarks. Subsequently, quarks and antiquarks from their respective pairs can merge, giving rise to the formation of a Higgs particle.
\vspace{10pt}
\begin{figure}[!h]
\centering
\begin{fmffile}{Topfusion}
\begin{fmfgraph*}(150,150)
  \fmfleft{i1,i2}
  \fmfright{o1,o2,o3}
  \fmf{gluon}{i1,v1}
  \fmf{gluon}{i2,v3}
  \fmf{fermion,label=$t$}{v1,o1}
  \fmf{fermion,label=$\bar t$}{o3,v3}
  \fmf{fermion,label=$\bar t$}{v2,v1}
  \fmf{fermion,label=$t$}{v3,v2}
  \fmf{dashes}{v2,o2}
  \fmflabel{$H$}{o2}
  \fmflabel{$g$}{i2}
  \fmflabel{$g$}{i1}
\end{fmfgraph*}
\end{fmffile}
\caption{Feynman diagram depicting the production of a Higgs boson through top quark fusion.}
\label{fig:Top-fusion}
\end{figure}
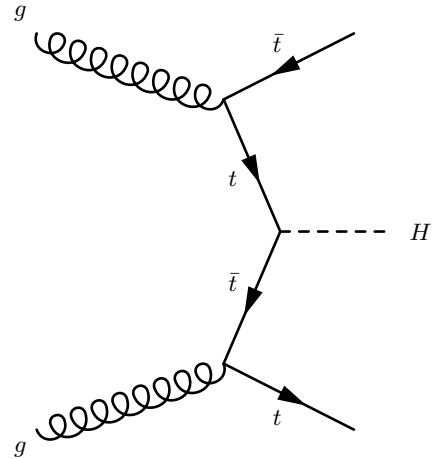
\subsection{Decay channels of Higgs boson}
\begin{figure}[!h]
  \centering
  \resizebox{\linewidth}{!}{\includegraphics{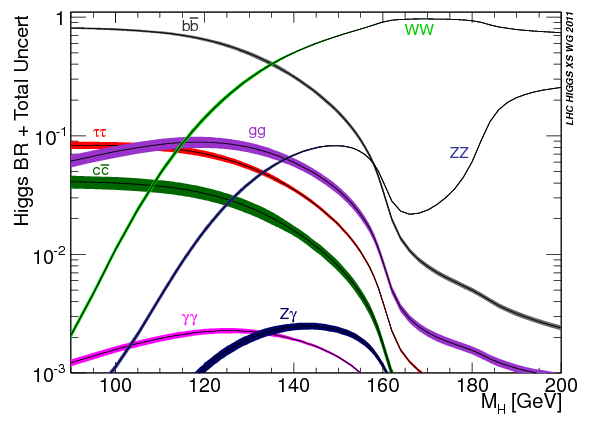}}
  \caption{Higgs branching ratios and their uncertainties for the low mass
range. \cite{lhc2011standard}}
  \label{fig:Higgs_branching_ratios}
\end{figure}

The Higgs boson, upon its creation, can undergo various decay processes, each with its own associated probabilities. The decay channels of the Higgs boson are determined by the particles it interacts with and the corresponding coupling strengths. The Standard Model predicts several possible decay modes for the Higgs boson, and their exploration is crucial in understanding its properties. The Higgs boson interacts with all the massive elementary particles in the Standard Model, resulting in various decay processes. Each of these processes has its own probability, known as the branching ratio, which represents the fraction of total decays following that specific process. The branching ratios are predicted by the Standard Model and are dependent on the mass of the Higgs boson (The predictions mentioned were made prior to the actual measurement of the Higgs boson's mass. However, once the measurement was conducted, the ratios were determined specifically for a mass value of $125 : GeV$). This relationship between the branching ratios and the Higgs mass is illustrated in Fig.(\ref{fig:Higgs_branching_ratios}).

One prominent decay mode is the decay of the Higgs boson into a pair of bottom quarks ($b\bar{b}$). This decay mode is particularly significant due to the Higgs boson's strong coupling to the bottom quark, which makes it one of the primary decay channels. However, it is also challenging to identify and measure accurately, due to the overwhelming background of other processes that produce bottom quarks.

Another important decay mode is the decay of the Higgs boson into a pair of W bosons ($WW$). This decay channel provides insights into the electroweak symmetry breaking mechanism and allows for studying the properties of the W boson. The subsequent decays of the W bosons can result in various final states, such as lepton-neutrino pairs or quark-antiquark pairs.
\vspace{10pt}
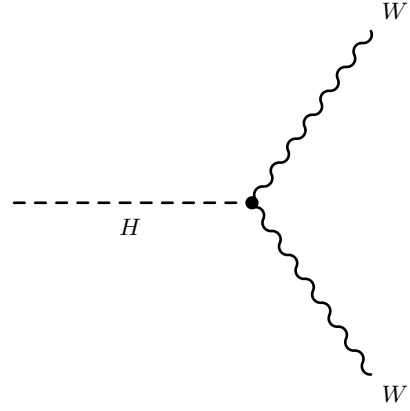
\begin{figure}[!h]
\centering
\begin{fmffile}{decay1}
\begin{fmfgraph*}(150,130)
  \fmfleft{i1}
  \fmfright{o1,o2}
  \fmf{photon}{v1,o1}
  \fmf{photon}{v1,o2}
  \fmf{dashes,label=$H$}{i1,v1}
  \fmfdot{v1}
  \fmflabel{$W$}{o2}
  \fmflabel{$W$}{o1}
\end{fmfgraph*}
\end{fmffile}
\caption{Feynman diagram of Higgs boson decays to $W$ bosons. \cite{aad2016measurements}}
\label{fig:decay1}
\end{figure}

Additionally, the Higgs boson can decay into a pair of $Z$ bosons ($ZZ$). This decay mode provides information about the Higgs boson's interaction with the $Z$ boson and offers a means to study the properties of the $Z$ boson. Similar to the $WW$ decay mode, the subsequent decays of the $Z$ bosons can produce different final states, including lepton pairs or quark-antiquark pairs.

Furthermore, the Higgs boson can decay into a pair of photons ($\gamma\gamma$). This decay mode is notable for its clean experimental signature, as photons are easy to detect accurately. Precise measurements of this decay channel help determine the Higgs boson's spin and its coupling to photons. The Standard Model (SM) predicts that the Higgs boson can also decay into a $Z$ boson and a photon which is similar to that of a decay into two photons. In these processes, the Higgs boson does not decay directly into these pairs of particles. Instead, the decays proceed via an intermediate "loop" of “virtual” particles that pop in and out of existence and cannot be directly detected. The  branching ratio for the decay $H\rightarrow Z\gamma$ is estimated to be around $0.15\%$. In a new study the two major experiments, CMS \cite{cms2022search} and ATLAS \cite{aad2020search}, at the Large Hadron Collider (LHC), have reported the first evidence of the Higgs boson decaying into a $Z$ boson and a photon. The measured signal rate shows a deviation of $1.9$ standard deviations from the prediction of the SM.
\vspace{10pt}
\begin{figure}[!h]
\begin{minipage}[b]{0.45\linewidth}
  \centering
  \begin{fmffile}{decay3}
    \begin{fmfgraph*}(100,130)
      \fmfleft{i1}
      \fmfright{o1,o2}
      \fmf{fermion,label.side=right,label=$t / b$}{v2,v1}
      \fmf{fermion,label.side=right,label=$ \bar t /\bar b$}{v3,v2}
      \fmf{fermion,label.side=right,label=$t / b$}{v1,v3}
      \fmf{dashes}{i1,v2}
      \fmf{photon}{v1,o1}
      \fmf{photon}{v3,o2}
      \fmfdot{v1,v2,v3}
      \fmflabel{$\gamma$}{o1}
      \fmflabel{$\gamma /Z$}{o2}
      \fmflabel{$H$}{i1}
    \end{fmfgraph*}
  \end{fmffile}
\end{minipage}
\hfill
\begin{minipage}[b]{0.45\linewidth}
  \centering
  \begin{fmffile}{decay33}
    \begin{fmfgraph*}(100,130)
      \fmfleft{i1}
      \fmfright{o1,o2}
      \fmf{photon,label.side=right,label=$W^+$}{v2,v1}
      \fmf{photon,label.side=right,label=$ W^-$}{v3,v2}
      \fmf{photon,label.side=right,label=$W^\pm$}{v1,v3}
      \fmf{dashes}{i1,v2}
      \fmf{photon}{v1,o1}
      \fmf{photon}{v3,o2}
      \fmfdot{v1,v2,v3}
      \fmflabel{$\gamma$}{o1}
      \fmflabel{$\gamma /Z$}{o2}
      \fmflabel{$H$}{i1}
    \end{fmfgraph*}
  \end{fmffile}
\end{minipage}
\caption{Feynman diagrams of Higgs boson decays to a pair of photons, or to a photon and a $Z$ boson. \cite{aad2016measurements}}
\label{fig:decay33}
\end{figure}
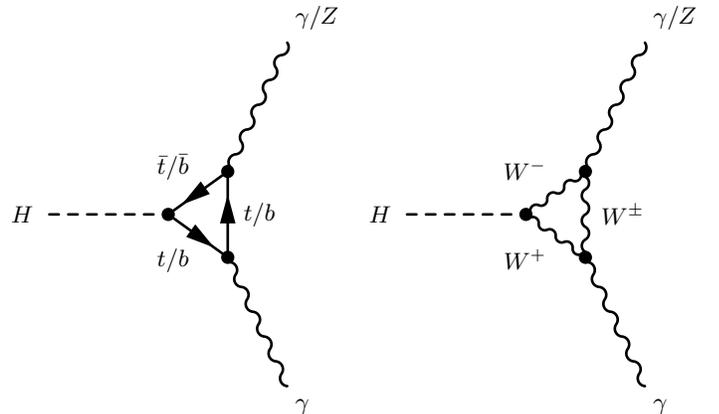

In addition to its primary decay modes, the Higgs boson manifests additional decay channels. These include its propensity to undergo transformations into tau leptons ($\tau^+\tau^-$), as well as the occurrence of more elusive decay modes involving muons, jets, or even the enigmatic phenomenon of missing energy. Such alternative decay channels offer invaluable opportunities for rigorous exploration and comprehensive analysis of the Higgs boson's intrinsic properties, as well as the potential manifestations of deviations from the well-established predictions of the Standard Model.
\vspace{10pt}
\begin{figure}[!h]
\centering
\begin{fmffile}{2}
\begin{fmfgraph*}(130,120)
  \fmfleft{i1}
  \fmfright{o1,o2}
  \fmf{fermion}{o1,v1}
  \fmf{fermion}{v1,o2}
  \fmf{dashes,label=$H$}{i1,v1}
  \fmfdot{v1}
  \fmflabel{$b, \tau^-, \mu^-$}{o2}
  \fmflabel{$\bar b, \tau^+, \mu^+$}{o1}
\end{fmfgraph*}
\end{fmffile}
\vspace{10pt}
\caption{Feynman diagrams of Higgs boson decays to to fermions. \cite{aad2016measurements}}
\label{fig:decay2}
\end{figure}
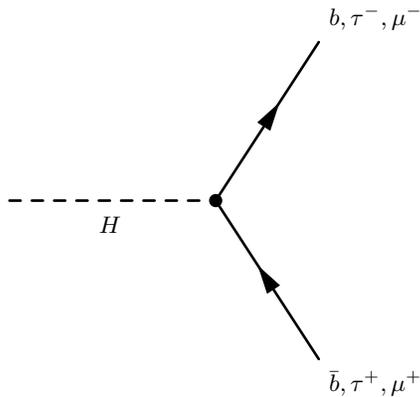
\section{\label{sec:level10} CONCLUSION}
In conclusion, our research has delved into the fascinating realm of the Higgs mechanism and its implications in our understanding of the fundamental building blocks of matter. We have explored the role of symmetries and spontaneous symmetry breaking, which provide crucial insights into the behavior of particles and the conservation laws that govern them. Through our investigation, we have discussed the foundations of the Standard Model, highlighting the $U(1)$ and $SU(2)$ gauge theories that underpin electromagnetism and the weak nuclear force, respectively. The unification of these forces into the electroweak theory has been a major milestone in the field of particle physics, highlighting the critical role of the Higgs mechanism in endowing mass to the $W$ and $Z$ bosons while unifying the electromagnetic and weak forces. This remarkable unification not only deepens our understanding of the fundamental nature of nature but also provides a comprehensive framework for investigating the intricate dynamics of particle interactions. Our exploration of the Higgs mechanism and the Higgs boson has provided significant insights into the fundamental nature of matter and the  mechanisms that govern our universe. While significant advancements have been achieved, many of unanswered queries and continuous investigations in the realm of Higgs physics propel us towards the prospect of unveiling further insights and breakthroughs.
 \appendix

\section{Klein–Gordon equation}
The Klein-Gordon equation is a fundamental equation in relativistic quantum mechanics that describes the behavior of particles with spin-zero, such as scalar bosons. Combining elements of special relativity and quantum mechanics, this equation provides insights into the wave-like nature and energy-momentum relationships of these particles.

In traditional non-relativistic quantum mechanics, the time-dependent Schrödinger equation describes the evolution of a particle's wave function over time.
\begin{equation*} i\frac{\partial \psi}{\partial t} +\frac{1}{2m}\nabla^2 \psi -V(\mathbf{r}, t)\psi=0. \end{equation*} 
However, when seeking a relativistic generalization, the equation needs to account for the relativistic energy-momentum relation. By quantizing the energy and momentum operators, the relativistic energy formula becomes
\begin{equation*}E=\sqrt {P^2+m^2},\end{equation*}
where $P$ represents momentum and $m$ denotes mass.
To derive the Klein-Gordon equation, Klein and Gordon started with the quantized version of the square of the relativistic energy formula \cite{bjorken1964relativistic}, which reads
\begin{equation*}{\bf p^2}+m^2={\bf E^2},\end{equation*}
inserting the quantum-mechanical operators for momentum and rearranging the terms, yielding the Klein-Gordon equation:
\begin{equation*}(\frac{\partial^2 }{\partial t^2}-\nabla^2+m^2 ) \psi=0.\end{equation*}
Rewriting the first two terms using the inverse of the Minkowski metric $diag(-1,1,1,1)$, which characterizes the geometry of spacetime, and writing the Einstein summation convention explicitly we get
\begin{equation*}-\eta^{\mu\nu}\partial_\mu \partial_\nu\equiv\sum_{\mu=0}^{3}\sum_{\nu=0}^{3} -\eta^{\mu\nu}\partial_\mu \partial_\nu=\frac{\partial^2 }{\partial t^2}-\nabla^2 .\end{equation*}
In this covariant notation, the d'Alembertian operator, often denoted as $\Box$ or Box, encapsulates the wave operator $\frac{\partial^2 }{\partial t^2}-\nabla^2$, resulting in the concise form of the Klein-Gordon equation:
\begin{equation}
(\Box +m^2)\psi=0,\label{ap:1}
\end{equation}
and its adjoint as
\begin{equation}
(\Box+m^2)\psi^*=0.\label{ap:2}
\end{equation}
This equation is foundational for understanding the dynamics and behavior of spin-zero particles in relativistic quantum mechanics.

The Klein-Gordon equation's covariant form provides a powerful mathematical framework for studying the wave-like properties and energy-momentum relationships of these particles. Its solutions reveal valuable insights into the fundamental nature of scalar bosons and their interactions.
By multiplying Eq.\eqref{ap:1} by $\psi$, and multiplying Eq.\eqref{ap:2} by $\psi^*$ and subtracting the result, we find that
\begin{equation*}\psi^* \Box \psi-\psi\Box\psi^*=0,\end{equation*}
\begin{equation*}\psi^* \left(\frac{\partial^2}{\partial t^2}-{\bf\nabla}^2\right) \psi-\psi\left(\frac{\partial^2}{\partial t^2}-{\bf \nabla}^2\right)\psi^*=0,\end{equation*}

\begin{equation*}\psi^* \frac{\partial^2\psi}{\partial t^2}-\psi^*{\bf \nabla}^2\psi -\psi \frac{\partial^2\psi^*}{\partial t^2}+\psi{\bf \nabla}^2\psi^*=0,\end{equation*}
\begin{equation*}\frac{\partial}{\partial t}\left(\psi^*\frac{\partial \psi}{\partial t}-\psi\frac{\partial \psi^*}{\partial t}\right)-{\bf \nabla}\cdot\left(\psi^*{\bf \nabla}\psi-\psi {\bf \nabla}\psi^*\right)=0. \end{equation*}
Comparing this to the continuity equation 
\begin{equation*}\frac{\partial \rho}{\partial t}+{\bf \nabla \cdot J}=0,\end{equation*}

we can define the probability current density as 
\begin{equation*}{\bf J}=\frac{1}{2 i m}\left(\psi^*{\bf \nabla}\psi-\psi {\bf \nabla}\psi^*\right),\end{equation*}
where $i$ is the imaginary unit and m is the particle's mass. By canceling out the imaginary components that result from taking the derivative of the complex conjugate of the wave function, the factor of $1/\left(2im\right)$ is used to verify that the probability current density J is a real-valued quantity.
and the current density as 
\begin{equation*}\rho=\frac{i}{2m}\left(\psi^*\frac{\partial \psi}{\partial t}-\psi\frac{\partial \psi^*}{\partial t}\right),\end{equation*}
where $m$ and $i$ appear in the same way as in the probability current density definition.

\section{Dirac equation}
Dirac \cite{dirac1928quantum} aimed to find a relativistic covariant wave equation in the form of Schrödinger's equation,
\begin{equation*}i\frac{\partial\psi}{\partial t}=\hat{H}\psi ,\end{equation*}
but with a positive definite probability density. The issue with the Klein-Gordon equation is that it is a second-order partial differential equation, which led to problematic negative probability solutions. In order to achieve an equation in a form similar to Schrödinger's equation, Dirac sought an equation that would be linear in both time and space. Thus, he proposed the Dirac equation in the following form:
\begin{gather}
i\frac{\partial \psi}{\partial t}=\Big[-i \left(\hat{\alpha}_1 \frac{\partial}{\partial x^1}+\hat{\alpha}_2 \frac{\partial}{\partial x^2}+\hat{\alpha}_3 \frac{\partial}{\partial x^3}\right)+\hat{\beta}m\Big]\psi\nonumber\\
=\Big[-i\sum ^N_{k=1}\hat{\alpha}_k \frac{\partial}{\partial x^k}+\hat{\beta}m\Big]\psi=\hat{H}_D \psi .\label{ap:3}
\end{gather}

This equation raised the suspicion that the coefficients should be matrices. Consequently, $\psi$ could not be a scalar, but instead had to be a column vector:
$$\psi=
\begin{pmatrix}
\psi_1(\bm{x},t)\\
\psi_2(\bm{x},t)\\
\vdots\\
\psi_N(\bm{x},t)\\
\end{pmatrix}.
$$
The probability density was defined as:
\begin{equation}
\rho=\psi^\dagger \psi= \sum_{i=1}^N \psi^\dagger_i \psi_i . \label{ap:4}
\end{equation}
As the aim was to construct a relativistic equation, it needed to satisfy certain criteria: first, it needed to be Lorentz covariant, ensuring that it preserved its form under Lorentz transformations; and second, the density defined in Eq.\eqref{ap:4} needed to satisfy a continuity equation, ensuring the conservation of probability.
For the first property of the equation, it suffices that each component of the spinor $\psi$ satisfies the Klein-Gordon equation Eq.\eqref{ap:1}, meaning (by rewriting of the Klein-Gordon equation):
\begin{equation}
-\frac{\partial^2 \psi}{\partial t^2}=(-\nabla^2+m^2 ) \psi=0. \label{ap:5}
\end{equation}
Multiplying Eq.\eqref{ap:3} by itself 
\begin{align*}
-\frac{\partial^2\psi}{\partial t^2}&=-\sum_{i,j=1}^3 \frac{\hat{\alpha}_i \hat{\alpha}_j+\hat{\alpha}_j \hat{\alpha}_i}{2}\frac{\partial\psi}{\partial x^i\partial x^j}\\
&\quad -im\sum_{i=1}^3 \left(\hat{\alpha}_i\hat{\beta}+\hat{\beta}\hat{\alpha}_i\right)\frac{\partial \psi}{\partial x^i\partial x^j}+\hat{\beta}^2m^2\psi
\end{align*}
Comparing coefficients between this expression and Eq.\eqref{ap:5} we immediately see that the following commutation relations must hold for the matrices $\hat{\alpha}_i$ and $\beta$:
\begin{equation}
\begin{aligned}
\hat{\alpha}_i\hat{\alpha}_j+\hat{\alpha}_j\hat{\alpha}_i&=\{\hat{\alpha}_i,\hat{\alpha}_j\}=2\delta_{ij}I_N, \\
\hat{\alpha}_i\hat{\beta}+\hat{\beta}\hat{\alpha}_i&=0, \\
\hat{\alpha}^2=\hat{\beta}^2&=I_N \label{ap:9}
\end{aligned}
\end{equation}
where $I_N$ is the $N-$dimensional identity matrix. Since the Hamiltonian $\hat{H}_D$ must be Hermitian, so it follows that $\hat{\alpha}_i$ and $\hat{\beta}$ must be Hermitian as well i.e., $\hat{\alpha}_i^\dagger=\hat{\alpha}_i$ and $\hat{\beta}^\dagger=\hat{\beta}$. Since $\hat{\alpha}^2=\hat{\beta}^2=I_N$, then the eigenvalues of these matrices can only have the values $\pm 1$. Multiplying the second equation by $\hat{\beta}$ from the right we get
\begin{equation*}\hat{\alpha}_i=-\hat{\beta}\hat{\alpha}_i\hat{\beta},\end{equation*}
and the trace of this equation is 
\begin{equation*}Tr\hat{\alpha}_i=-Tr\hat{\beta}\hat{\alpha}_i\hat{\beta}=-Tr\hat{\beta}^2\hat{\alpha}_i=-Tr\hat{\alpha}_i .\end{equation*}
Thus, $Tr\hat{\alpha}_i=0.$ Now, since the trace of a matrix equals to the sum of its eigenvalues, therefore there must be an even number of eigenvalues which equals to $\pm 1$ in order to insure that $Tr\hat{\alpha}_i=Tr\hat{\beta}=0$. The only condition that is remains to be satisfied is the first one,there are only three anticommuting matrices which satisfies this condition (Pauli matrices), which take to form:
$$\hat{\sigma}_1=
\begin{pmatrix}
0&&1\\
1&&0
\end{pmatrix}, \quad \hat{\sigma}_2=
\begin{pmatrix}
0&&-i\\
i&&0
\end{pmatrix},\quad \hat{\sigma}_3=
\begin{pmatrix}
1&&0\\
0&&-1
\end{pmatrix}.
$$
Dirac chose the simplest condition that satisfies the equation that is the dimension of the matrices to equal $4$. Therefore, the following choice for $\hat{\alpha}_i$ and $\hat{\beta}$ satisfies all the conditions
$$\hat{\alpha}_i=
\begin{pmatrix}
0&&\hat{\sigma}_i\\
\hat{\sigma}_i&&0
\end{pmatrix},\quad \hat{\beta}=
\begin{pmatrix}
I_2&&0\\
0&&I_2
\end{pmatrix}.
$$
Now, we want to ensure that the continuity equation is satisfied, first, we construct the four-current density as follows.
By multiplying Eq.\eqref{ap:3} by $\psi^\dagger=(\psi_1^*,\psi_2^*,\psi_3^*,\psi_4^*)$ and get
\begin{equation}
i\psi^\dagger \frac{\partial \psi}{\partial t}=-i\sum ^N_{k=1}\psi^\dagger\hat{\alpha}_k \frac{\partial \psi}{\partial x^k}+\psi^\dagger\hat{\beta}m\psi . \label{ap:6}
\end{equation}
Then, multiplying the conjugate of Eq.\eqref{ap:3} by $\psi$ from the right, and noting that $\hat{\alpha}_i^\dagger=\hat{\alpha}_i$ and $\hat{\beta}^\dagger=\hat{\beta}$ we get
\begin{equation}
-i\psi\frac{\partial \psi^\dagger}{\partial t}=i\sum ^N_{k=1}\psi\hat{\alpha}_k \frac{\partial \psi^\dagger}{\partial x^k}+\psi\hat{\beta}m\psi^\dagger , \label{ap:7}
\end{equation}
then, subtracting Eq.\eqref{ap:7} from Eq.\eqref{ap:6}, we get
\begin{align*}
i\psi^\dagger \frac{\partial \psi}{\partial t}+i\psi\frac{\partial \psi^\dagger}{\partial t}&=-i\sum ^N_{k=1}\psi^\dagger\hat{\alpha}_k \frac{\partial \psi}{\partial x^k}+\psi^\dagger\hat{\beta}m\psi\\
&\quad  -i\sum ^N_{k=1}\psi\hat{\alpha}_k \frac{\partial \psi^\dagger}{\partial x^k}-\psi\hat{\beta}m\psi^\dagger ,
\end{align*}
which is equivalent to 
$$\frac{\partial}{\partial t}\left(\psi^\dagger \psi\right)+\sum ^N_{k=1}\frac{\partial }{\partial x^k}\left(\psi^\dagger \hat{\alpha}_k\psi\right)=0,
$$
which has the form of the continuity equation 
\begin{equation*}\frac{\partial \rho}{\partial t}+\nabla \cdot J=0,\end{equation*}
where 
\begin{equation*}
    \rho=\psi^\dagger \psi=\sum_{i=1}^4\psi_i^* \psi .
\end{equation*}
The expected conservation law yields as usual by integrating over the space: 
$$\frac{\partial}{\partial t}\int d^3x \quad \psi^\dagger \psi =0.
$$

So, the probability density $\rho$ is positive, definite, and conserved. 
For the theory to be consistent with the principles of relativity, the equations of motion must exhibit Lorentz covariance. The preferred notation in the realm of relativity is to represent all four-vectors using the symbol $x$. Consider two inertial frames, $S$ and $S'$, moving at a relative velocity of $v$ with respect to each other. Suppose an event occurs simultaneously in both frames, and its coordinates in $S$ are given by $\{ct,x^1,x^2,x^3,x^4\}$, while its coordinates in $S'$ are given by $\{ct',x^{'1},x^{'2},x^{'3},x^{'4}\}$. Given a wave function $\psi$ in the frame of reference $S$, the Lorentz transformation must enable us to compute the corresponding wave function $\psi'$ in the frame of reference $S'$. To satisfy the principle of Lorentz covariance, the transformation must preserve the form of the Dirac equation, such that each wave function satisfies the same equation in both reference frames. we have to change the notations a bit, Beginning with Eq. \eqref{ap:3} given by
\begin{equation*}\left(i\frac{\partial}{\partial t}+i \sum ^3_{k=1}\hat{\alpha}_k \frac{\partial}{\partial x^k}-\hat{\beta}m \right)\psi=0,\end{equation*}
we can obtain a new expression by multiplying the equation by $\hat{\beta}$ from the left, yielding
\begin{equation*}\hat{\beta}\left(i\frac{\partial}{\partial t}+i \sum ^3_{k=1}\hat{\beta}\hat{\alpha}_k \frac{\partial}{\partial x^k}-m \right)\psi=0.\end{equation*}
Then, by defining $\hat{\beta}=\gamma^0$ and $\gamma^\mu=\hat{\beta}\hat{\alpha}_k$, for $\mu=1,2,3$, we can rewrite the equation as
\begin{equation*}i\left(\gamma^0\frac{\partial}{\partial x^0}+\gamma^1\frac{\partial}{\partial x^1}+\gamma^2\frac{\partial}{\partial x^2}+\gamma^3\frac{\partial}{\partial x^3}\right)\psi-m\psi=0.\end{equation*}
The anticommutation relations Eq.\eqref{ap:9} now read
\begin{equation*}\gamma^\mu \gamma^\nu +\gamma^\nu \gamma^\mu=2g^{\mu \nu }I_4 ,\end{equation*}
here, $\gamma^\mu$ are unitary, and anti-Hermitian. On the other hand, $\gamma^0$ is unitary and Hermitian. We also write an explicit representation for these matrices:
$$\gamma^\mu=
\begin{pmatrix}
0&&\hat{\sigma}^\mu\\
-\hat{\sigma}^\mu&&0
\end{pmatrix},\quad \gamma^0=
\begin{pmatrix}
I_2&&0\\
0&&-I_2
\end{pmatrix}.
$$
Then we can write the Dirac equation in the famous form
\begin{equation*}\left(i\gamma^\mu \partial_\mu-m\right)\psi=0,\end{equation*}
which can be verified that it is Lorentz invariant provided that $\gamma^\mu$ obeys the commutation rules Eq.\eqref{ap:9}. 

The Lagrangian $\mathcal{L}$ and the field equations are generally equivalent, although the Lagrangian arguably seems more fundamental, we can obtain the field equations given the Lagrangian, but inverting the process is less straightforward.
The Lagrangian that describes the field of single free fermion with mass $m$ is 
\begin{equation}\mathcal{L}(\mathrm{x})=\bar{\psi}(\mathrm{x})\left(i\gamma^\mu\partial_\mu -m \right)\psi(\mathrm{x}),\label{ap:10}\end{equation}
then, differentiating the Lagrangian with respect to $\psi(x)$, yields
\begin{equation*}\frac{\partial \mathcal{L}}{\partial \psi (\mathrm{x})}=-\bar{\psi}(\mathrm{x})m, \end{equation*}
and differentiating the Lagrangian with respect to $\partial_\mu \psi(x)$, yields
\begin{equation*}\frac{\partial \mathcal{L}}{\partial\left(\partial_\mu \psi (\mathrm{x})\right)}=i\gamma^\mu\bar{\psi}(\mathrm{x}).\end{equation*}
Therefore, the Euler-Lagrange equation, in this case, gives 
\begin{equation*}\frac{\partial \mathcal{L}}{\partial \psi}-\partial_\mu \frac{\partial \mathcal{L}}{\partial(\partial_\mu \psi)}=-\bar{\psi}(\mathrm{x})m -i\gamma^\mu\partial_\mu\bar{\psi}(\mathrm{x})=0.\end{equation*}
Thus,
\begin{equation*}\bar{\psi}(\mathrm{x})\left(i\gamma^\mu\partial_\mu+m \right)=0,\end{equation*}
where, $\bar{\psi}=\psi^\dagger \gamma^0$, then
\begin{equation*}\psi^\dagger(\mathrm{x}) \gamma^0)\left(i\gamma^\mu\partial_\mu+m \right)=0.\end{equation*}
Taking the Hermitian conjugate of this equation, we obtain,
\begin{equation*}\gamma^0\left(-i\left(\gamma^0 \gamma^\mu \gamma^0\right)\partial_mu+m\right)\psi=0,\end{equation*}
where $\gamma^{\mu \dagger}=\gamma^0 \gamma^\mu \gamma^0$, and $\left(\gamma^0\right)^2=1$. Therefore
\begin{equation*}\left(-i\left( \gamma^\mu \gamma^0\right)\partial_\mu+m\gamma^0\right)\psi=0.\end{equation*}
Finally, we have
\begin{equation*}\left(i\gamma^\mu \partial_\mu-m\right)\psi=0.\end{equation*}
Which is Dirac equation. In other words, we see that the Lagrangian density denoted as Eq.\eqref{ap:10} yields the Dirac equation for a free fermionic field as its Euler-Lagrange equation.
\begin{widetext}
\onecolumngrid
\begin{acknowledgments}
   I would like to express my deepest appreciation and gratitude to my supervisor, Dr. Alaa Abd El-hady, for his valuable suggestions, exceptional guidance, mentorship, and support throughout the course of this review article.
\end{acknowledgments}
\twocolumngrid
\end{widetext}

\bibliography{references}
\end{document}